\documentclass[preprint,aps,floatfix,nofootinbib,groupedaddress,superscriptaddress]{revtex4}
\usepackage{float}
\usepackage{amsfonts}
\usepackage{amssymb}
\usepackage{savesym}
\usepackage{amsmath}
\savesymbol{iint}
\usepackage{txfonts}
\restoresymbol{TXF}{iint}
\usepackage{ctable}

\usepackage{epsfig}
\usepackage{textcomp}
\usepackage[page]{appendix}
\usepackage{bm}

\begin{document}

\title{Entanglement of polar molecules in pendular states}

\author{Qi Wei}
\affiliation{Department of Physics, Texas A $\&$ M University,
College Station, TX 77843, USA}\affiliation{Department of Chemistry,
Purdue University, West Lafayette, IN 47907, USA}
\author{Sabre Kais}
\affiliation{Department of Chemistry, Purdue University, West
Lafayette, IN 47907, USA}
\author{Bretislav Friedrich}
\affiliation{ Fritz-Haber-Institut der Max-Planck-Gesellschaft,
Faradayweg 4-6, D-14195 Berlin, Germany}
\author{ Dudley Herschbach \footnote{ Corresponding email: dherschbach@yahoo.com}} \affiliation{Department of Physics,
Texas A $\&$ M University, College Station, TX 77843, USA}

\begin{abstract}
In proposals for quantum computers using arrays of trapped ultracold
polar molecules as qubits, a strong external field with appreciable
gradient is imposed in order to prevent quenching of the dipole
moments by rotation and to distinguish among the qubit sites. That
field induces the molecular dipoles to undergo pendular
oscillations, which markedly affect the qubit states and the
dipole-dipole interaction. We evaluate entanglement of the pendular
qubit states for two linear dipoles, characterized by pairwise
concurrence, as a function of the molecular dipole moment and
rotational constant, strengths of the external field and the
dipole-dipole coupling, and ambient temperature. We also evaluate a
key frequency shift, $\triangle\omega$, produced by the
dipole-dipole interaction. Under conditions envisioned for the
proposed quantum computers, both the concurrence and
$\triangle\omega$ become very small for the ground eigenstate. In
principle, such weak entanglement can be sufficient for operation of
logic gates, provided the resolution is high enough to detect the
$\triangle\omega$ shift unambiguously. In practice, however, for
many candidate polar molecules it appears a challenging task to
attain adequate resolution. Simple approximate formulas fitted to
our numerical results are provided from which the concurrence and
$\triangle\omega$ shift can be obtained in terms of unitless reduced
variables.

\end{abstract}

\maketitle

\section{Introduction}

Since the original proposal by DeMille~\cite{Demille}, arrays of
ultracold ($<$ 1 mK) polar molecules have come to be considered
among the most promising platforms to implement a quantum computer
~\cite{carr,Book2009,Friedrich,Lee,Kotochigova,yelin,Micheli,Charron,kuz,ni,lics,YelinDeMille}.
His proposal describes a complete scheme for quantum computing using
as qubits  the dipole moments of diatomic molecules, trapped in a
one-dimensional optical lattice, partially oriented in an external
electric field, and coupled by the dipole-dipole interaction.   The
qubit states are individually addressable because the field has an
appreciable gradient so the Stark effect is different for each
location in the array.

A subsequent proposal has advocated coupling polar molecules into a
quantum circuit using superconducting wires~\cite{wallraff}. Such
capacitive, electrodynamic coupling to transmission line resonators
is analogous to coupling to Rydberg atoms and Cooper pair
boxes~\cite{sorensen,andre}.   The molecular qubits are entangled
via the coupling to the transmission lines rather than direct
dipole-dipole interactions. Again, addressability of the qubits is
achieved via the Stark effect by means of local gating of an
electrostatic field.

Entanglement is a major ingredient in most quantum computation
algorithms. It is among the defining features of quantum mechanics,
with no classical analog~\cite{Book,Book2,Siegfried}. A pure state
of a pair of quantum systems is said to be entangled if its
wavefunction cannot be factored into a product of wavefunctions of
the individual partners. For example, the singlet state of two
spin-$\frac{1}{2}$ particles,
$\frac{1}{\sqrt{2}}(|\uparrow\downarrow\rangle-|\downarrow\uparrow\rangle)$
is entangled. A mixed state is entangled if it cannot be represented
as a mixture of factorizable pure states. The allure of quantum
information processing has recently motivated studies of
entanglement for a variety of potential qubit
systems\cite{entg-kais,Vedral,Osterloh,Amico,Su,Dur,Cincio,Charron,Lee,Kotochigova,Osenda1,Osenda2,Osenda3,Osenda4,Xu,Wei}.
These include one-dimensional arrays of localized spins, coupled
through exchange interactions and subject to an external magnetic
field~\cite{entg-kais} and analogous treatments of trapped electric
dipoles coupled by dipole-dipole interactions~\cite{Wei}.

However, the previous studies of entanglement of electric dipoles
have not adequately considered how the external electric field,
integral to current designs for quantum computers using polar
molecules, affects both the qubit states and the dipole-dipole
interaction. For the simplest case of a $^1\Sigma$ diatomic
molecule, the qubit eigenstates resulting from the Stark effect are
linear combinations of spherical harmonics, with coefficients that
depend markedly on the field strength.   These are appropriately
termed {\it pendular} states ~\cite{Friedrich2}, or field-dressed
states \cite{Ticknor}. In such states, the orientation of the dipole
moment has a broad angular range (not solely along or opposed to the
field direction as are spins in a magnetic field). Likewise, the
dipole-dipole interaction for molecules in pendular states is much
different than that for dipoles in the absence of an external field.

Here we evaluate entanglement, as measured by pairwise concurrence,
for the prototype case of two diatomic polar molecules in pendular
states, ultracold and trapped in distinct optical lattice sites. The
molecules are represented as identical rigid dipoles, undergoing
angular oscillations, a fixed distance apart and subject either to a
different or to the same external electric field.  We examine the
dependence of the concurrence on three dimensionless variables.  The
first governs the energy and intrinsic angular shape of the qubits
(when the dipole-dipole interaction is switched off). It is
$\mu${\Large $\mathbf{ \varepsilonup}$}/$B$, the ratio of the Stark
energy (magnitude of permanent dipole moment times electric field
strength) to the rotational constant (proportional to inverse of
molecular moment of inertia).  The second variable governs the
magnitude of the dipole-dipole coupling.  It is $\Omega/B$, with
$\Omega$ = ($\mu^2/r^3$), the square of the permanent dipole moment
divided by the cube of the separation distance. The third variable,
$k_BT/B$, is the ratio of thermal energy (Boltzmann constant times
Kelvin temperature) to the rotational constant.

We also examine an aspect related to but distinct from entanglement.
The operation of a quantum gate~\cite{Jones} such as CNOT requires
that manipulation of one qubit (target) depends on the state of
another qubit (control). This is characterized by the shift,
$\triangle\omega$, in the frequency for transition between the
target qubit states when the control qubit state is changed. The
shift $\triangle\omega$, which is due to the dipole-dipole
interaction, must be kept {\it smaller} than the differences
required to distinguish among addresses of qubit sites. Under
conditions envisaged in the proposed designs
\cite{Demille,carr,Book2009,Friedrich,Lee,YelinDeMille} for quantum
computing with trapped polar molecules, $\Omega/B < 10^{-4}$, and
for the ground eigenstate both the entanglement and frequency shift
$\triangle\omega$ become very small. For CNOT and other operations,
entanglement needs to be large, but can be induced dynamically, so
need not be appreciable in the ground eigenstate. Yet a small
$\triangle\omega$ shift can only suffice if the resolution is high
enough to detect the shift unambiguously. From estimates of the line
widths of transitions between the pendular qubit states, we find it
an open question whether adequate resolution can be obtained for
typical candidate diatomic molecules.

\section{ENTANGLEMENT FOR TWO DIPOLES IN PENDULAR STATES}

\subsection{Hamiltonian terms and pendular qubit states}

The Hamiltonian for a single trapped linear polar molecule in an
external electric field is
\begin{equation}
{\bf H}=\frac{p^2}{2m} + V_{trap}({\bf r}) + B\bf{J}^2 -
\boldsymbol{\mu}\cdot \text{\Large \boldsymbol{$\varepsilonup$}}
\end{equation}
where the molecule, with mass $m$, rotational constant $B$ and
body-fixed dipole moment $\mu$, has translational kinetic energy
$p^2/2m$, potential energy $V_{trap}$ within the trapping field, and
rotational energy $B\bf{J}^2$ as well as interaction energy
$\boldsymbol{\mu}\cdot${\Large\boldsymbol{$\varepsilonup$}} with the
external field {\Large\boldsymbol{$\varepsilonup$}}. In the trapping
well, at ultracold temperatures, the translational motion of the
molecule is quite modest and very nearly harmonic; $p^2/2m +
V_{trap}({\bf r})$ thus is nearly constant and can be omitted from
the Hamiltonian.  There remains the rotational kinetic energy and
Stark interaction,
\begin{equation}
{\bf H_S} = B\bf{J}^2 - {\bf \mu}\text{\Large $\mathbf{
\varepsilonup}$}\text{cos}\theta
\end{equation}
which represent a spherical pendulum with $\theta$ the polar angle
between the molecular axis and the field direction.   Figure
\ref{Figure1}(a) displays the lowest few pendular
eigenenergies~\cite{Hughes} for a $^1\Sigma$ diatomic (or linear)
molecule, as functions of $\mu${\Large $\varepsilonup$}/$B$. These
are labeled with the familiar quantum numbers $\tilde{J}$, M that
specify the field-free rotational states. However, $\tilde{J}$ wears
a tilde to indicate it is no longer a good quantum number since the
Stark interaction mixes the rotational states, whereas M (denoting
the projection of the ${\mathbf J}$-vector on the field direction)
remains good as long as azimuthal symmetry about {\Large
$\varepsilonup$} is maintained. As proposed by DeMille, the qubit
states $|0\rangle$ and $|1\rangle$ are chosen as the lowest M = 0
pendular states, with $\tilde{J}$ = 0 and 1, respectively. These are
superpositions of $Y_{j,0}$ spherical harmonics,
\begin{equation}
|0\rangle=\sum_{j}a_jY_{j,0}(\theta,\varphi),\;\;\;\;\;\;\;\;\;\;
|1\rangle=\sum_{j}b_jY_{j,0}(\theta,\varphi)
\end{equation}
Figure \ref{Figure2} plots the coefficients as functions of
$\mu${\Large $\varepsilonup$}/$B$. Figure \ref{Figure3} displays the
angular distributions of the pendular qubit states. For $|0\rangle$
the distribution is unimodal and as $\mu${\Large
$\varepsilonup$}/$B$ increases the dipole orientation increasingly
favors the direction of the {\Large $\varepsilonup$}-field (at
$\theta$ = $0^o$).  For $|1\rangle$ the distribution is bimodal
because, with M = 0, the dipole is rotating perpendicular to the
${\mathbf J}$-vector, which is perpendicular to the field direction.
For {\Large $\varepsilonup$} = 0, the dipole orientation is equally
probable in the hemispheres toward ($\theta < 90^o$) or opposite
($\theta
> 90^o$) to the field direction. As $\mu${\Large $\varepsilonup$}/$B$
increases, the pinwheeling dipole favors the opposite hemisphere
because there its motion is slowed because the Stark interaction
becomes unfavorable. However, when $\mu${\Large $\varepsilonup$}/$B$
becomes large enough, pinwheeling is inhibited and converted into
pendular libration about the field direction, so the dipole
orientation shifts to favor the toward hemisphere.

Adding a second trapped polar molecule, identical to the first but
distance $r_{12}$ apart, introduces in addition to its pendular term
the dipole-dipole coupling interaction,
\begin{equation}
V_{d-d} =\frac{\bm{\mu}_1\cdot\bm{\mu}_2-3(\bm{\mu}_1 \cdot {\bf
n})(\bm{\mu}_2 \cdot {\bf n})}{|{\bf r}_1-{\bf r}_2|^3}
\label{coupling}
\end{equation}
Here {\bf n} denotes a unit vector along ${\bf r}_{12}$.  In the
presence of an external field, it becomes appropriate to express
$V_{d-d}$ in terms of angles related to the field direction.  As
shown in Appendix A, the result after averaging over azimuthal
angles (that for M = 0 states are uniformly distributed) reduces to
\begin{equation}
V_{d-d}=\Omega(1-3\text{cos}^2\alpha)\text{cos}\theta_1\text{cos}\theta_2
\label{coupling2}
\end{equation}
where $\Omega=\mu^2/r_{12}^3$, the angle $\alpha$ is between the
${\bf r}_{12}$ vector and the field direction and polar angles
$\theta_1$ and $\theta_2$ are between the ${\bm \mu}_1$ and ${\bm
\mu}_2$ dipoles and the field direction. Until later (Sec. IV), we
consider the external field magnitude and direction to be the same
at the sites of both the polar molecules.

\subsection{Entanglement measured by pairwise concurrence}

We will deal with the entanglement of formation,
$\mathfrak{E}(\rho)$, which characterizes the amount of entanglement
needed in order to prepare a state described by a density matrix,
$\rho$. (Henceforth, we term $\mathfrak{E}(\rho)$ just
"entanglement", for short.) Wootters \cite{Wootters,Hill} has shown
that $\mathfrak{E}(\rho)$ for a general state of two qubits can be
quantified by the pairwise $\it concurrence$, $C(\rho)$, which
ranges between zero and unity. The relation can be written as
\begin{equation}
\mathfrak{E}(\rho) = \xi(C(\rho))
\end{equation}
where $\xi$ is given by
\begin{equation}
\xi(C) = h\left ( \frac{1+\sqrt{1-C^2}}{2} \right )
\end{equation}
with $h(x)=-x \text{log}_2 x - (1 - x) \text{log}_2(1- x)$. The
function $\xi(C)$ increases monotonically between zero and unity as
$C$ varies from 0 to unity. The concurrence is given by
\begin{equation}
C(\rho) = max
\left\{0,\sqrt{\lambda_1}-\sqrt{\lambda_2}-\sqrt{\lambda_3}-\sqrt{\lambda_4}
\right\} \label{concurrence}
\end{equation}
where the $\lambda_i$'s are the eigenvalues, in decreasing order, of
the non-Hermitian matrix $\rho\tilde{\rho}$, where $\tilde{\rho}$ is
the density matrix of the spin-flipped state, defined as
\begin{equation}
\tilde{\rho} =
(\sigma_y\otimes\sigma_y)\rho^*(\sigma_y\otimes\sigma_y)
\label{rhotilde}
\end{equation}
with $\rho^*$ the complex conjugate of $\rho$ and $\sigma_y$ a Pauli
matrix. The parent density matrix $\rho$ is taken in the basis
formed by combining the pendular qubit states; for a pair of
two-level particles, this comprises the four state vectors
$\{|00\rangle,|01\rangle,|10\rangle,|11\rangle\}$.

In order to evaluate thermal entanglement, we need a temperature
dependent density matrix, $\rho = \text{exp}(-\beta H)/Z(T)$, with
$\beta = 1/k_BT$ and $Z(T)$ the partition function
\begin{equation}
Z(T) = tr[\text{exp}(-\beta H)] = \sum_{i}g_ie^{-\beta E_i}
\end{equation}
with $E_i$ the $i^{th}$ eigenvalue and $g_i$ its degeneracy. Hence
the density matrix can be written as
\begin{equation}
\rho(T) = \frac{1}{Z}\sum_{i}^{N}e^{-\beta
E_i}|\Psi_i\rangle\langle\Psi_i|
\end{equation}
where $|\Psi_i\rangle$ is the $i^{th}$ eigenfunction. From the
density matrix $\rho(T)$, we can obtain the reduced density matrix
for any pair of dipoles and thence evaluate the concurrence at any
temperature.

\section{Concurrence of two dipoles in pendular states}

\begin{figure}[t]
\begin{center}
\includegraphics[width=1.0\textwidth]{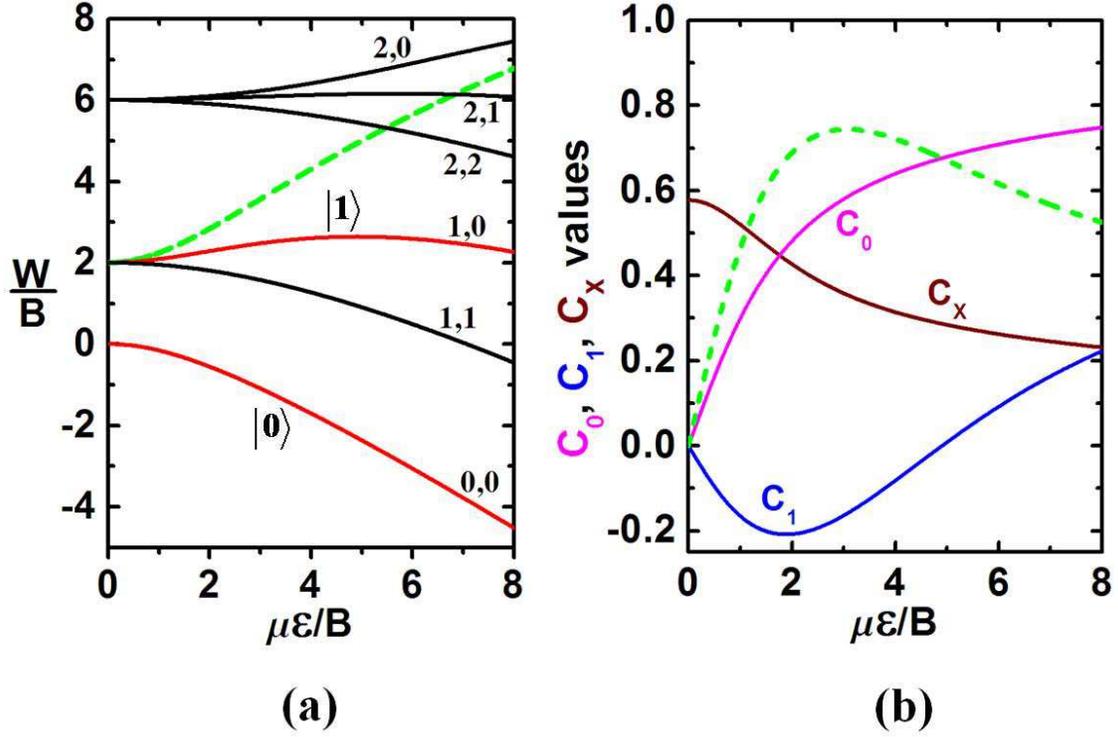}
\end{center}
\caption{(Color online) Stark states for a polar diatomic molecule
in a $^1\Sigma$ electronic state \cite{Hughes}, as functions of
$\mu${\Large $\mathbf{ \varepsilonup}$}$/B$, with $\bf \mu$ the
permanent dipole moment, {\Large $\varepsilonup$} the field
strength, $B$ the rotational constant.  (a) Eigenenergies, W, and
(b) Matrix elements of orientation cosines; see Eq.(14). States used
as qubits (red curves) are labeled $|0\rangle$ and $|1\rangle$. In
the field-free limit, $|0\rangle$ correlates with the $J = 0$, $M_J
= 0$ and $|1\rangle$ with the $J = 1$, $M_J = 0$ rotational states.
Dashed curve (green) in (a) shows energy for transition between
qubit states, $\triangle W = W_1 - W_0$; that in (b) shows $C_0 -
C_1$, difference between effective dipole moments, projections of
the molecular dipole on the field direction for pendular states
$|0\rangle$ and $|1\rangle$.} \label{Figure1}
\end{figure}

\begin{figure}[t]
\begin{center}
\includegraphics[width=1.00\textwidth]{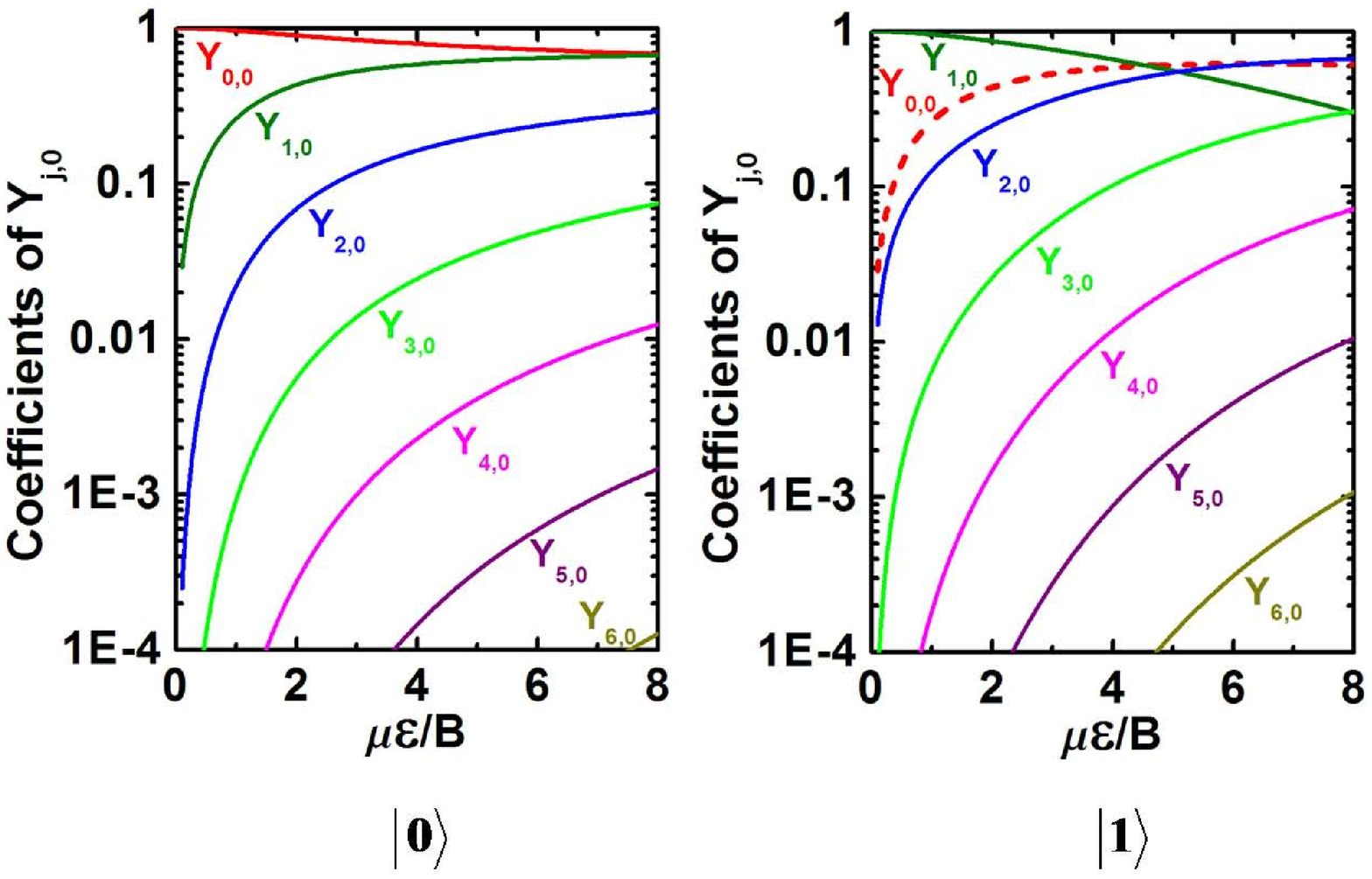}
\end{center}
\caption{(Color online) Coefficients of sums of spherical harmonics
for pendular states $|0\rangle$ and $|1\rangle$, see Eq.(3). Dashed
curve for $|1\rangle$ indicates the coefficient of $Y_{0,0}$ is
negative.} \label{Figure2}
\end{figure}

\begin{figure}[t]
\begin{center}
\includegraphics[width=1.00\textwidth]{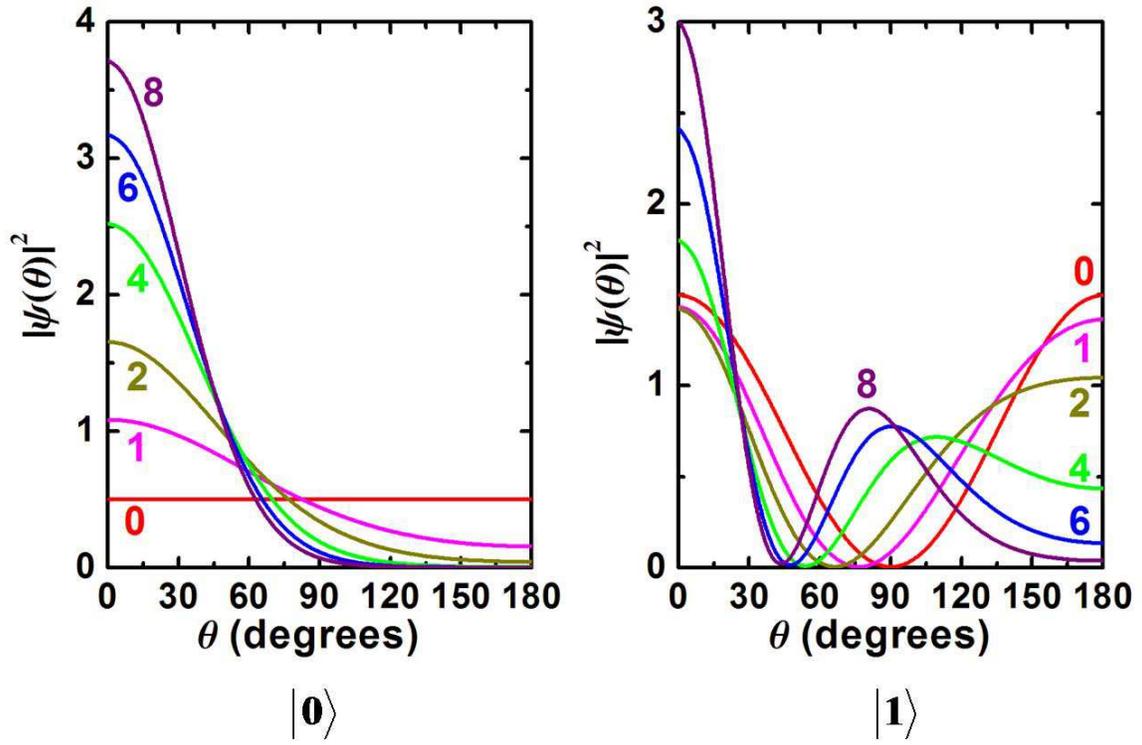}
\end{center}
\caption{(Color online) Angular distributions of the $|0\rangle$ and
$|1\rangle$ pendular states for values of $\mu${\Large
$\varepsilonup$}$/B$ between 0 and 8.} \label{Figure3}
\end{figure}

We illustrate the calculation of pairwise concurrence for $N=2$
dipoles. The Hamiltonian, $H_{S1}$ + $H_{S2}$ + $V_{d-d}$, when set
up in a basis of the qubit pendular states,
$\{|00\rangle,|01\rangle,|10\rangle,|11\rangle\}$, takes the form,
\begin{equation}
{\bf H_{S1} + H_{S2}}=\left(\begin{array}{cccc}
W_0+W_0^{\prime}  & 0       & 0       & 0   \\
 0    & W_0+W_1^{\prime} & 0       & 0   \\
 0    & 0       & W_1+W_0^{\prime} & 0  \\
 0    & 0       & 0       & W_1+W_1^{\prime}
\end{array} \right)
\label{Hs-matrix}
\end{equation}
\begin{equation}
{\bf V_{d-d}}=\Omega(1-3\text{cos}^2\alpha)
\left(\begin{array}{cccc}
C_0C_0^{\prime}  & C_0C_X^{\prime} & C_XC_0^{\prime} & C_XC_X^{\prime}   \\
 C_0C_X^{\prime} & C_0C_1^{\prime} & C_XC_X^{\prime} & C_XC_1^{\prime}   \\
 C_XC_0^{\prime} & C_XC_X^{\prime} & C_1C_0^{\prime} & C_1C_X^{\prime}  \\
 C_XC_X^{\prime} & C_XC_1^{\prime} & C_1C_X^{\prime} & C_1C_1^{\prime}
\end{array} \right)
\label{Hdd-matrix}
\end{equation}
where $W_0$ and $W_1$ are the eigenenergies of the pendular qubit
states $|0\rangle$, and $|1\rangle$, in the absence of the
dipole-dipole interaction. Primes attached to quantities for the
second dipole indicate that the external field magnitude may differ
at its site (although, as noted above, we postpone evaluating that
case until Sec. IV). In $V_{d-d}$ the basis qubit states are linked
by matrix elements containing  factors arising from the orientation
cosines in Eq. (\ref{coupling2}); these are
\begin{equation}
C_0=\langle 0|\text{cos}\theta|0\rangle;\;\;\;\;C_X=\langle
0|\text{cos}\theta|1\rangle;\;\;\;\;C_1=\langle
1|\text{cos}\theta|1\rangle
\end{equation}
$C_0$ and $C_1$ are the expectation values of cos$\theta$ in the
pendular states $|0\rangle$ and $|1\rangle$, respectively, so
represent for those states the effective dipole moment projections
displayed in Fig.\ref{Figure1}(b). $C_X$ corresponds to an exchange
interaction or transition dipole moment between the qubit states.
Both the Stark eigenenergies $W_i$ and the dipole-dipole elements
$C_k$ are functions of $\mu${\Large $\varepsilonup$}/$B$. As seen in
Fig.\ref{Figure1}(b), as $\mu${\Large $\varepsilonup$}/$B$ is
increased $C_0$ becomes increasingly positive, whereas $C_1$ is
increasingly negative until about $\mu${\Large $\varepsilonup$}/$B$
= 2, then climbs to zero at about $\mu${\Large $\varepsilonup$}/$B$
= 4.9 and thereafter is increasingly positive. The range
$\mu${\Large $\varepsilonup$}/$B$ = 2 to 4 is recommended for the
proposed quantum computer designs \cite{Demille,andre}; within that
range, the difference in the effective dipole moments of the qubits,
$|C_0 - C_1|$, varies only modestly.

If the dipole-dipole interaction is omitted ($\Omega$ = 0), the
eigenvectors of $H_{S1}$ + $H_{S2}$ are simply $\Psi_1 =
|00\rangle$, $\Psi_2 = 2^{-1/2}(|10\rangle - |01\rangle)$, $\Psi_3 =
2^{-1/2}(|10\rangle + |01\rangle)$, $\Psi_4 = |11\rangle$,
corresponding to the eigenenergies of Eq.(12). For $\Psi_1$ and
$\Psi_4$, which are obviously nonentangled states, the concurrence
is zero. For $\Psi_2$ and $\Psi_3$, which exemplify fully entangled
states, the concurrence is unity; these are termed Bell states
\cite{Book}.

When the dipole-dipole coupling is included, an analytical solution
to obtain eigenstates is only feasible when the external field is
switched off.   As shown in Appendix B, in that limit analytical
results can be obtained for each step in evaluating the concurrence,
both for the four individual eigenstates and their combination in
the thermal concurrence.  As seen in Fig. 1, for $\mu${\Large
$\varepsilonup$}/$B$ = 0, the energy terms in Eq.(12) involve merely
$W_0$ = 0 and $W_1$ = 2$B$. In the $V_{d-d}$ matrix of Eq.(13), the
cosine matrix elements $C_0$ and $C_1$ vanish and $C_X$ =
$3^{-1/2}$; thus, the only nonzero elements occur along the
antidiagonal and (for $\alpha$ = $90^o$) are just $\Omega C_X^2$.
The results for this zero-field limit prove useful in interpreting
those for the general pendular case.

The limits with $\Omega$ = 0 and/or $\mu${\Large
$\varepsilonup$}/$B$ = 0 motivate setting up the Hamiltonian of
Eqs.(12) and (13), for the (unprimed) case with the same external
field at both dipole sites, using a basis of Bell states:
\begin{equation}
\frac{|11\rangle + |00\rangle}{\sqrt{2}},\;\;\; \frac{|11\rangle -
|00\rangle}{\sqrt{2}},\;\;\;\frac{|10\rangle +
|01\rangle}{\sqrt{2}},\;\;\;\frac{|10\rangle -
|01\rangle}{\sqrt{2}},
\end{equation}
In this basis, the Hamiltonian becomes
\begin{equation}
{\bf H_{S1} + H_{S2}}=\left(\begin{array}{cccc}
W_+   & W_-     & 0       & 0   \\
W_-   & W_+     & 0       & 0   \\
 0    & 0       & W_+     & 0  \\
 0    & 0       & 0       & W_+
\end{array} \right)
\label{Hs-matrixbell}
\end{equation}
\begin{equation}
{\bf V_{d-d}}=\Omega(1-3\text{cos}^2\alpha)
\left(\begin{array}{cccc}
\hat{A}_+     & \hat{B}     & \hat{C}_+       & 0   \\
\hat{B}       & \hat{A}_-   & \hat{C}_-       & 0   \\
\hat{C}_+     & \hat{C}_-   & \hat{D}_+       & 0  \\
 0            & 0           & 0               & \hat{D}_-
\end{array} \right)
\label{Hdd-matrixbell}
\end{equation}
Where $W_{\pm} = W_1 \pm W_0$ and $\hat{A}_{\pm} = \frac{1}{2}(C_1^2
+ C_0^2) \pm C_X^2$, $\hat{B} = \frac{1}{2}(C_1^2 - C_0^2)$,
$\hat{C}_{\pm} = C_X(C_1 \pm C_0)$, $\hat{D}_{\pm} = C_1C_0 \pm
C_X^2$. This makes explicit a consequence of the symmetry between
the (unprimed) sites \cite{Thankreviewer}. In the Bell basis, the
Hamiltonian factors, with the state $2^{-1/2}(|10\rangle -
|01\rangle)$ in a $1 \times 1$ block, so that state remains
maximally entangled regardless of the value of $\mu${\Large
$\varepsilonup$}/$B$ or $\Omega/B$.

\begin{figure}[t]
\begin{center}
\includegraphics[width=1.00\textwidth]{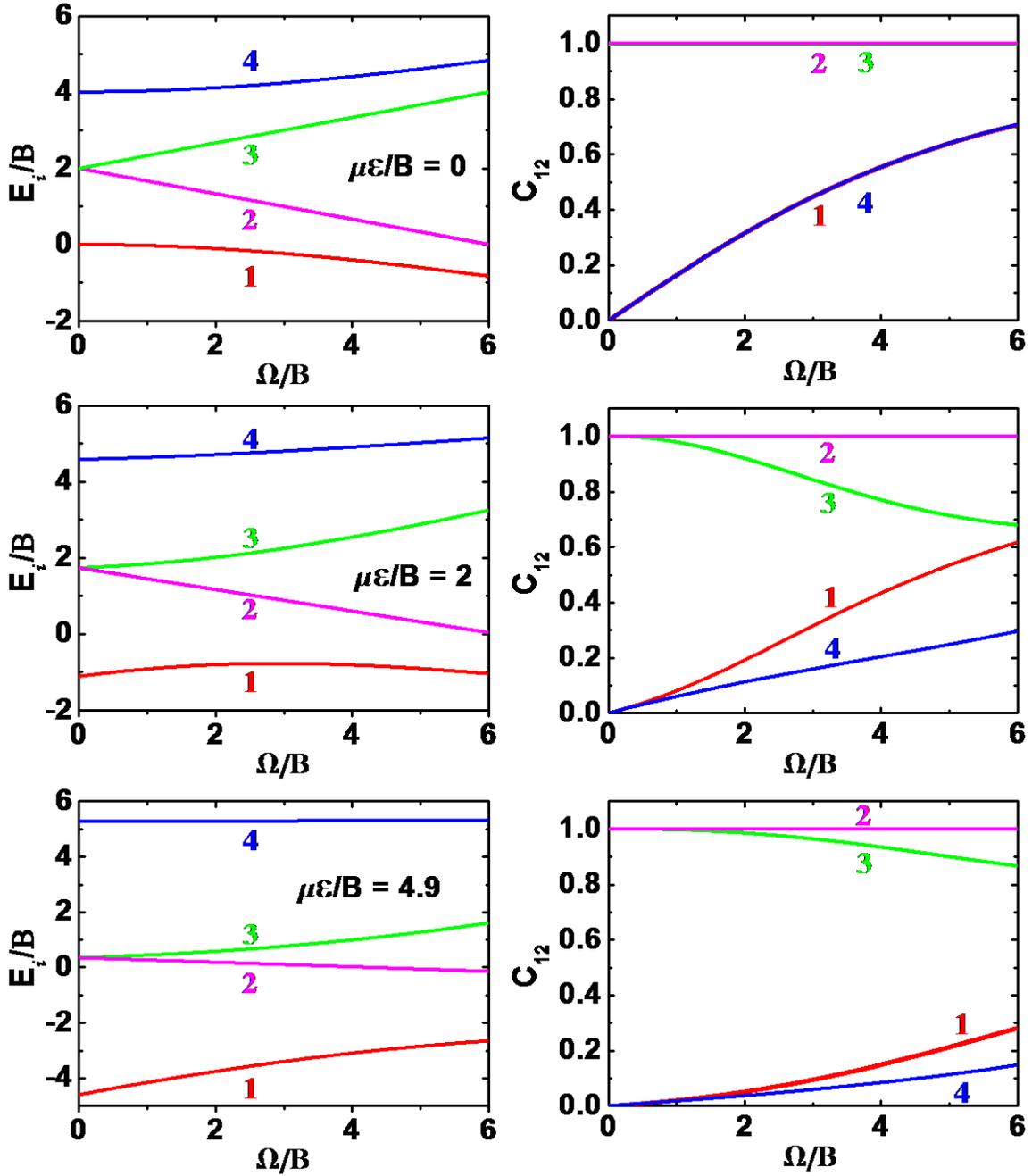}
\end{center}
\caption{(Color online) Eigenenergies, numbered 1, 2, 3, 4 in order
of increasing energy, and pairwise concurrences of the eigenstates
for two dipoles as a function of the dipole-dipole coupling constant
$\Omega/B$ for three values of the reduced electric field strength,
$\mu${\Large $\varepsilonup$}$/B$ = 0, 2 and 4.9.} \label{Figure4}
\end{figure}

Figure \ref{Figure4} plots, for $\mu${\Large $\varepsilonup$}/$B$ =
0, 2 and 4.9, the eigenenergy and pairwise concurrence versus
$\Omega/B$ = 0 to 6 for the four eigenstates of the two-dipole
system. The eigenstates are numbered from 1 to 4 in order of
increasing energy. For $\mu${\Large $\varepsilonup$}/$B$ = 0, both
eigenstates 2 and 3 are Bell states, with eigenenergies $E_i/B$ = $2
- (\Omega/6B)$ and $2 + (\Omega/6B)$, respectively; eigenstates 1
and 4 are also entangled (much more weakly) by the dipole-dipole
interaction, with eigenenenergies that shift downwards and upwards
nonlinearly with increasing $\Omega/B$, respectively. For
$\mu${\Large $\varepsilonup$}/$B$ $>$ 0, the concurrences increase
with $\Omega/B$ for eigenstates 1 and 4, and decrease for eigenstate
3. By virtue of the symmetry imposed factorization noted above,
eigenstate 2 retains the same Bell form despite the Stark and
dipole-dipole interactions which affect its energy, and its
concurrence is always unity. For small $\Omega/B << 1$, eigenstate 3
also becomes independent of the dipole-dipole interaction and
coincides with eigenstate 2 in both energy and concurrence.   For
$\mu${\Large $\varepsilonup$}/$B$ = 4.9, as seen in Fig. l(b), the
$C_1=\langle 1|\text{cos}\theta|1\rangle$ factor that appears in
seven of the matrix elements in Eq. (13) vanishes. Consequently, the
energy of eigenstate 4 then becomes independent of the dipole-dipole
interaction, although its wavefunction and concurrence do not.

\begin{figure}[t]
\begin{center}
\includegraphics[width=1.00\textwidth]{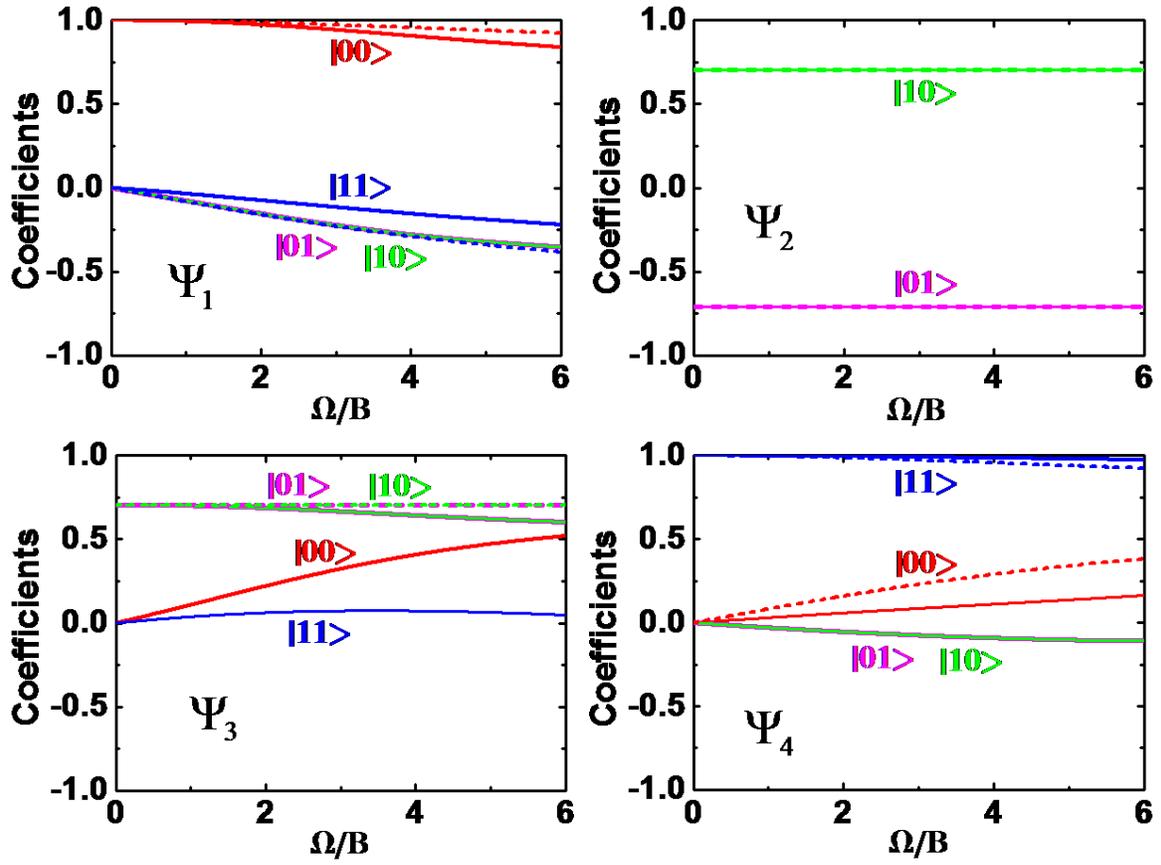}
\end{center}
\caption{(Color online) Eigenvectors of the four eigenstates for two
dipoles as a function of the dipole-dipole coupling constant
$\Omega/B$ = 0 to 6 for $\mu${\Large $\varepsilonup$}$/B$ = 0
(dashed curves) and 2 (solid curves).} \label{Figure5}
\end{figure}

Figure \ref{Figure5} shows, for $\mu${\Large $\varepsilonup$}/$B$ =
0 and 2, how the contributions of the basis states to each of the
eigenstates vary with the strength of the dipole-dipole interaction.
This illustrates that for $\Omega/B << 1$ the eigenstates rapidly
approach those for $\Omega$ = 0. Indeed, we find that for $\Omega/B
< 0.04$ the concurrences for eigenstates 1 and 4, which rapidly
become the same, are proportional to $\Omega/B$ within better than
1\%. Thus,
\begin{equation}
C_{12} = K(x)[\Omega/B]\label{C12}
\end{equation}
where the proportionality factor $K(x)$ is a function of $x$ =
$\mu${\Large $\varepsilonup$}/$B$. At the zero-field limit, $K(0) =
1/6$. In Appendix C we describe a numerical analysis that provided
an accurate approximate formula,
\begin{equation}
K(x) = A_1 + \frac{A_2}{1+\text{exp}[(x-x_0)/\triangle x]}
\label{kc}
\end{equation}
This is plotted in Fig.6 and values of the four parameters are
listed in Appendix C.

Figure \ref{Figure7} displays for $\Omega/B$ = 0.1, 1 and 6 the
eigenenergies and concurrences versus $\mu${\Large
$\varepsilonup$}/$B$ from 0 to 8 for the four eigenstates. As the
dipole-dipole interaction increases 60-fold over this range, its
effect on the eigenstate energies is relatively modest, whereas the
concurrences change markedly, in response to variations in
eigenvector compositions such as illustrated in Fig. \ref{Figure5}.

Figure \ref{Figure8} gives a contour plot of the thermal pairwise
concurrence derived from Eq.(11) as a function of $\Omega/B$ and
$k_BT/B$. It pertains to $\mu${\Large $\varepsilonup$}/$B$ = 3; we
found that normalizing the thermal concurrence to its value for T =
0 and $\Omega/B$ = 1 removed most of the variation with $\mu${\Large
$\varepsilonup$}/$B$ from such contour plots. For T = 0, the thermal
concurrence coincides with that for the ground state, eigenstate
$\Psi_1$. However, as $k_BT/B$ increases, the thermal concurrence
decreases and is always smaller than the ground-state concurrence.
This may seem odd, because Eq.(11) specifies a shift in population
that reduces the contribution from the gound state, while bringing
in contributions from the excited states.  The eigenstates 2 and 3
then populated have large concurrence, so increasing temperature
might be expected to make the net thermal concurrence become larger
than for the ground-state, rather than smaller.  The source of this
behavior is indicated by the analytic solution obtained in Appendix
B for the zero-field limit,
\begin{equation}
C_{12}(\text{T}) = C_{12}(1)P_1 - C_{12}(2)P_2 - C_{12}(3)P_3 -
C_{12}(4)P_4 \label{C12-Thermal}
\end{equation}
where $P_i = (1/Z)\text{exp}(-E_i/k_BT)$ with $Z(T) =
\sum_{i}\text{exp}(-E_i/k_BT)$. This shows that the excited states
indeed reduce the thermal concurrence, an effect traceable to Eq.(8)
and which persists even for large $\mu${\Large $\varepsilonup$}/$B$.

Another striking aspect of Fig. 8 is that the concurrence vanishes
along and outside a particular contour.   That contour defines
mutually dependent maximum values of $k_BT/B$ and minimum values of
$\Omega/B$ required to obtain nonzero concurrence. When $\Omega/B <<
1$, we find that a modified form of Eq.(\ref{C12}) represents the
thermal concurrence,
\begin{equation}
C_{12}(\text{T}) = \text{max}\{0, K(x)[y - y_0(x,z)]\}
\label{C12thermal2}
\end{equation}
Here $x$ = $\mu${\Large $\varepsilonup$}/$B$; $y$ = $\Omega/B$; and
$z = k_BT/B$ is the scaled temperature. Fig. \ref{Figure9} gives a
contour plot of $y_0 = \Omega_c/B$, the critical dipole-dipole
coupling required for nonzero concurrence. Some further details are
included in Appendix B.

The original proposal by DeMille and kindred papers on quantum
computing with trapped polar molecules
\cite{Demille,carr,Lee,yelin,kuz,YelinDeMille,andre} discuss for
several examples the range of experimental conditions deemed
suitable and acceptable. For trap temperatures of the order of a
microkelvin or below, the typical values of $k_BT/B$ are a few times
$10^{-6}$, so indicate that only ground-state entanglement would be
significant. The external field strengths considered are typically a
few $kV/cm$. The spacing between optical lattice sites, $r =
\lambda/2$, is half the optical lattice wavelength.  The optimal
choice of $\lambda$ ranges between 1 to 0.3 microns, depending on
electronic transition frequencies of the molecules to be trapped
\cite{YelinDeMille}. From these parameters and molecular data,
values of $\Omega/B$ are small; we find for a dozen potential
candidate molecules values ranging between $4 \times 10^{-6}$ (for
KCs) to $2 \times 10^{-4}$ (for CsI). A favorite candidate is SrO
($\mu = 8.9 D$, $B = 0.33$ $cm^{-1}$, $\lambda = 1$ micron), for
which $\Omega/B \sim 10^{-5}$. In that regime, the concurrence is
simply proportional to $\Omega/B$, so can be easily evaluated from
Eq.(\ref{C12}) and/or (\ref{C12thermal2}) without use of the rather
elaborate prescription outlined in Eqs.(6-14).

\section{Frequency shift for two coupled dipoles in pendular states}

In the region $\Omega/B < 10^{-4}$, the concurrence of the ground
eigenstate is very small, typically $< 10^{-5}$. However, such
meager entanglement in eigenstates can still be adequate for quantum
computing, as demonstrated with NMR versions of quantum computers
\cite{Jones2}. The key aspect is that although entanglement needs to
be large for some quantum computing algorithms, it need not be
appreciable or even present in the ground eigenstate of the system;
it can be induced dynamically during operation of the computer
\cite{Zhou}. Here, for the polar molecule case, we consider this
aspect. We also evaluate an eigenstate property, a small frequency
shift, distinct from but related to the pairwise concurrence, that
is important for quantum computing.

The need for selective excitation in operation of quantum logic
gates \cite{Jones,Seth,Cory} is an essential feature. Taking the
2-qubit CNOT gate as an example, its operation requires that
manipulation of one qubit (target) is perceptively affected by the
state of the other qubit (control).  In our case, the qubits are
pendular states that can be accessed by microwave transitions, which
offer high spectral resolution.  As resolution has a crucial role,
we now suppose the external field differs enough at the two dipole
sites (denoted unprimed and primed) to supply distinct addresses for
the sites ({\it cf.} Fig. 1(a), green dashed curve).

Since $\Omega/B$ is so small, we first omit the dipole-dipole
interaction and, as illustrated in Fig. \ref{Figure10}, consider
transitions among the pendular eigenstates of Eq.(12). Although in
this limit the ground-state concurrence is zero, as seen in
Eq.(\ref{C12}), it is possible to generate states of large
concurrence by use of resonant pulses \cite{Zhou,Thanks}. Start by
applying a pulse resonant with the transition denoted $\omega_1$,
between $|00\rangle$ and $|01\rangle$, which has energy $W'_1 -
W'_0$.  Note that $\omega_1$ needs to be well-resolved from the
transition $\omega_3$, between $|00\rangle$ and $|10\rangle$, which
has energy $W_1 - W_0$. The separation thus comes from the different
values of the external field at the two sites (plus a dipole-dipole
contribution, in higher order).  The requisite field strength
difference, {\Large $\varepsilonup$}$'$ - {\Large $\varepsilonup$},
can be readily determined from another approximation formula,
\begin{equation}
(W_1 - W_0)/B = A_1 +\frac{A_2}{1 + \left (x/x_0\right )^p}
\label{eqW1-W0}
\end{equation}
by comparing results for $x$ = $\mu${\Large $\varepsilonup$}/$B$ and
$x'$ = $\mu'${\Large $\varepsilonup$}$'$/$B$; the accurate fit
obtained (better than 1\% except near $x=0$) is displayed in Fig.
\ref{Figure11} and the four parameters in Eq.(\ref{eqW1-W0}) are
given in Appendix C. The amplitude and duration of the $\omega_1$
pulse can be adjusted to make it a $\pi/2$ pulse, which will put the
system in the state $2^{-1/2}(|00\rangle + |01\rangle)$.

Next, to complete the CNOT gate, apply a pulse resonant with the
transition $\omega_2$ between $|01\rangle$ and $|11\rangle$. This
needs to be well-resolved from transition $\omega_3$ between
$|00\rangle$ and $|10\rangle$. However, in our initial
approximation, both $\omega_2$ and $\omega_3$ have the same
transition energy, $W_1 - W_0$. Hence, weak as it is, the
dipole-dipole interaction is seen to have an essential role: to
introduce a frequency shift, $\triangle\omega = \omega_3 -
\omega_2$, adequate for unambiguous resolution. If that is
fulfilled, the amplitude and duration of the $\omega_2$ pulse can be
adjusted to make it a $\pi$ pulse. Thereby the system will be put in
the state $2^{-1/2}(|00\rangle + |11\rangle)$. This result of a CNOT
gate is to first approximation a Bell state (aside from small
corrections of order $\Omega/B$), so its concurrence will be near
unity. It is not an eigenstate, so will evolve with time but in
principle would remain nearly fully entangled until degraded by
other interactions.

\renewcommand{\theequation}{23\alph{equation}}
\setcounter{equation}{0}  

If now the dipole-dipole terms from Eq.(13) are included to first
order, we obtain
\begin{equation}
\omega_1 = \langle 01|\hat{H}|01\rangle - \langle
00|\hat{H}|00\rangle = W'_1 - W'_0 + \Omega_{\alpha}C_0(C'_1-C'_0)
\end{equation}
\begin{equation}
\omega_2 = \langle 11|\hat{H}|11\rangle - \langle
01|\hat{H}|01\rangle = W_1 - W_0 + \Omega_{\alpha}C'_1(C_1-C_0)
\end{equation}
\begin{equation}
\omega_3 = \langle 10|\hat{H}|10\rangle - \langle
00|\hat{H}|00\rangle = W_1 - W_0 + \Omega_{\alpha}C'_0(C_1-C_0)
\end{equation}
\begin{equation}
\omega_4 = \langle 11|\hat{H}|11\rangle - \langle
10|\hat{H}|10\rangle = W'_1 - W'_0 + \Omega_{\alpha}C_1(C'_1-C'_0)
\end{equation}
where $\Omega_{\alpha} = \Omega(1-3\text{cos}^2\alpha)$. Thus, the
key frequency shift is given by
\renewcommand{\theequation}{\arabic{equation}}
\setcounter{equation}{23}  
\begin{equation}
\triangle\omega = \omega_3 - \omega_2 = \omega_4 - \omega_1 =
\Omega_{\alpha}(C_1-C_0)(C'_1-C'_0)
\end{equation}
For given $\Omega_{\alpha}$, the frequency shift $\triangle\omega$
depends only on $x$ and $x'$, which determine at the respective
sites the difference in the effective dipole moment projections
$C_0$ and $C_1$ along the external electric field, specified in
Eq.(14). To provide a convenient means to evaluate Eqs.(23) and (24)
we again fitted our numerical results to obtain accurate
approximation formulas,
\begin{equation}
C_0(x) = A_1 +\frac{A_2}{1 + \left (x/x_0\right )^p} \label{eqC0}
\end{equation}
\begin{equation}
C_1(x) = A_0 + \frac{A_1}{1+\text{exp}[(x-x_1)/\triangle
x_1]}+\frac{A_2}{1+\text{exp}[-(x-x_2)/\triangle x_2]} \label{eqC1}
\end{equation}
These functions are plotted in Fig. \ref{Figure11}, together with
$C_0 - C_1$, and the fitted parameters are given in Appendix C.

Since for small $\Omega/B$, both the concurrence and
$\triangle\omega$ are proportional to $\Omega/B$, the frequency
shift provides an equivalent measure of entanglement. When the
{\Large $\varepsilonup$}-fields differ at the two sites, Eq.(18)
still provides a very accurate approximation for $C_{12}(x, x')$,
merely by replacing the proportionality factor by the geometric
mean, $[K(x)K(x')]^{1/2}$. The concurrence (which involves $C_X$,
the exchange interaction term) is in principle different from
$\triangle\omega$ but both have about the same magnitude. The
frequency shift is much more relevant for quantum computing, because
$\triangle\omega$ is directly involved in the CNOT gate.

Also important, in addition to the pulse shapes which affect the
population transfers, are the durations of the resonant pulses
required to resolve $\omega_1$ and $\omega_2$ from $\omega_3$; these
must satisfy $\tau_{31} >> 1/|\omega_3 - \omega_1|$ and $\tau_{32}
>> 1/|\omega_3 - \omega_2|$. For $\tau_{31}$ the lower bound usually
can be made very low, permitting a short pulse duration. This holds
because $\triangle${\Large $\varepsilonup$} as well as dipole-dipole
terms contribute to $|\omega_3 - \omega_1|$, which thus can be made
large by choice of the {\Large $\varepsilonup$}-field gradient,
regardless of whether $\Omega_{\alpha}$ is extremely small. In
contrast, for $\tau_{32}$ the separation $\triangle \omega =
|\omega_3 - \omega_2|$ depends only on the dipole-dipole
interaction.  The smaller $\triangle \omega$ is, the longer the
$\omega_2$ pulse duration has to be in order to complete the CNOT
operation. Although larger $\triangle\omega$ allows a shorter pulse
duration, $\triangle\omega$ must not be so large that it becomes
comparable to or larger than the addressing shift produced by
$\triangle${\Large $\varepsilonup$}, thereby thwarting correct
identification of the qubits.

\begin{table}[H]
\centering \caption{Illustrative CNOT Gate Implementation$^a$.}
\begin{minipage}{15cm}
\begin{tabular}{cccccccc}
\hline\hline
 $\mu${\Large
$\varepsilonup$}/$B$ &\;\;\; $x=1$ \;\;& \;\; $x'=1.01$ \;\;& \;\; $x'=1.10$ &\;\;\;\; & $x=3$ \;\;& \;\; $x'=3.03$ \;\;& \;\;$x'=3.30$ \\
\hline\let\thefootnote\relax\footnotetext{$^a$For $\Omega_{\alpha}/B
= 10^{-5}$; $x$ = $\mu${\Large $\varepsilonup$}/$B$, $x'$ =
$\mu${\Large $\varepsilonup$}$'$/$B$. As the quantities in the
lowest three rows are functions of both $x$ and $x'$, their values
are listed in the $x'$ columns. There E-n denotes a factor of
$10^{-n}$.}
 $(W_1-W_0)/B$  &\;\;\; 2.2709 \;\;& \;\; 2.2759 \;\;& \;\; 2.3218 &\;\;\;\; & 3.5614 \;\;& \;\; 3.5831 \;\;& \;\;3.7789  \\
 $C_0$  &\;\;\; 0.30165 \;\;& \;\; 0.30404 \;\;& \;\; 0.32487 &\;\;\;\; & 0.57922 \;\;& \;\; 0.58149 \;\;& \;\;0.60051  \\
 $C_1$  &\;\;\; -0.16467 \;\;& \;\; -0.16573 \;\;& \;\; -0.17461 &\;\;\;\; & -0.16362 \;\;& \;\; -0.16150 \;\;& \;\;-0.14115  \\
 $C_0-C_1$  &\;\;\; 0.46632 \;\;& \;\; 0.46977 \;\;& \;\; 0.49948 &\;\;\;\; & 0.74284 \;\;& \;\; 0.74298 \;\;& \;\;0.74165  \\
\hline
 $(\omega_1-\omega_3)/B$  &\;\;\;  &\;\;\; 4.99E-3 \;\;&\;\; 5.09E-2 &\;\;\;\; &\;\;\;\;  \;\;& \;\;2.17E-2 \;\;& \;\;2.17E-1  \\
 $\triangle\omega/B$  &\;\;\;  &\;\;\; 2.19E-6 \;\;&\;\; 2.33E-6 &\;\;\;\; &\;\;\;\;  \;\;& \;\;5.52E-6 \;\;&\;\; 5.51E-6  \\
 $C_{12}$  &\;\;\;  &\;\;\; 1.20E-6 \;\;&\;\; 1.17E-6 &\;\;\;\; &\;\;\;\;  \;\;& \;\;3.57E-7 \;\;&\;\; 3.34E-7  \\
\hline
\end{tabular}
\end{minipage} \label{tableexample}
\end{table}

Table I provides specific numbers pertaining to the SrO example.
From Sec.III, we take $\Omega_{\alpha}/B=10^{-5}$. As representative
{\Large $\varepsilonup$}-field values, we use $x$ = 1 and 3 for site
1 and and take $x'$ higher by 1\% or 10\% for site 2.   From
Eqs.(23), the transition frequencies $\omega_1 = \omega_4$ and
$\omega_2 = \omega_3$ (in units of $B$) to 5 or 6 significant
figures. The frequency difference that must be resolvable for the
first step of the CNOT operation, $\omega_1 - \omega_3$, is
approximately just $\triangle\triangle W = (W'_1- W'_0) - (W_1 -
W_0)$. From Fig. \ref{Figure11}, this is seen to grow about linearly
with both $x$ and $x' - x$. The values in Table I (third row from
bottom) range from $> 10^{-3}$ to $> 10^{-1}$ (in units of $B$). To
accommodate more dipole qubits, it may be desired to make much
smaller the {\Large $\varepsilonup$}-field differences between
sites; steps with $\triangle x$ = 0.01\% were proposed by DeMille
\cite{Demille}. That might encounter engineering limitations, but in
principle the proportionally smaller $\omega_1 - \omega_3$
difference could still be readily resolved.  For the second step of
the CONT operation the crucial frequency shift, $\triangle\omega =
\omega_1 - \omega_3$ varies only modestly with $x$ and practically
not at all with $x' - x$. The values of $\triangle\omega/B$ in Table
I (second row from bottom) range between 2 and $6 \times 10^{-6}$;
in frequency units, this range is 20 to 60 kHz. Smaller still are
the corresponding values of the concurrence (bottom row); also
insensitive to $x' - x$ but, in accord with Fig. 6, varying more
rapidly with $x$.

Figure \ref{Figure12} exhibits for both the $\triangle\omega$ shift
and concurrence the variation with $\alpha$, the angle between the
direction of the electric field and the axis between the dipoles.
This dependence enters via the factor ($1 - 3\text{cos}^2\alpha$) in
the dipole-dipole interaction, Eq.(5), which emerges directly in the
$\triangle\omega$ shift, Eq.(24), and by a more complex route
propagates into the concurrence, via Eq.(13). Tilting the field
direction to make $\alpha = 54.73^\circ$, the "magic angle",
provides a simple means to shut off the entanglement. That is a
useful option, awkward to attain in other ways
\cite{yelin,YelinDeMille}.

\section{Conclusions and prospects}

In this study, our chief aim has been to examine entanglement of
polar molecules by the dipole-dipole interaction and subject to an
external electric field, for the prototype case of two diatomic or
linear $^1\sum$ molecules. This required use of qubits that are
pendular states comprised of sums of spherical harmonics. We focused
on the pairwise concurrence and its dependence on three unitless
reduced variables, involving the dipole moments, field strength,
rotational constant, dipole-dipole coupling and temperature. We have
considered a wide range of the parameters, to map general features
of the concurrence. However, for conditions envisioned for proposed
quantum computers, the dipole-dipole coupling is weak ($\Omega/B$
typically of order $10^{-4}$ to $10^{-6}$) and the concurrence
becomes very small ($< 10^{-5}$). For that weak coupling realm, we
found the $\triangle\omega$ frequency shift provides an equivalent
measure of entanglement, directly related to observable properties
and hence preferable to the concurrence. We also obtained for both
the $\triangle\omega$ shift and concurrence in the weak realm simple
explicit formulas in terms of the reduced variables.

For quantum computing a crucial issue is whether $\triangle\omega$
is large enough to enable the $\omega_2$ transition to be reliably
distinguished from $\omega_3$ (and, equivalently, $\omega_1$ from
$\omega_4$). For typical candidate polar molecules, this requires
resolving transitions separated by only tens of kHz. That would not
be feasible in conventional molecular spectroscopy. Under ordinary
gas phase conditions, transitions between molecular rotational or
pendular states have line widths of the order of a few 100 kHz
\cite{Townes}. For ultracold molecules trapped in an optical
lattice, line widths may be much narrower. Collisional broadening is
eliminated and at microkelvin temperatures Doppler broadening is
also quenched (as trap conditions are in the Lamb-Dicke regime). It
is encouraging that for ultracold atoms extremely narrow line widths
have been attained by exploiting "magic" optical trapping conditions
that are expected to be at least in part applicable to molecules
\cite{Kotochigova2}. At present, however, no data have been reported
on line widths for rotational transitions of ultracold molecules
trapped in an optical lattice and subject to an external electric
field. In view of the small size of $\triangle\omega$, it is
important to obtain such data to assess the resolution attainable,
since motion within the traps, coupling to lattice fields, and
inhomogeneity of the external field may introduce appreciable line
broadening.

We have sought to glean pertinent evidence from electric resonance
spectroscopy of molecular beams, as the beams are collision free and
transitions are observed in an external electric field ("Rabi
C-field"). For BaO, both $\triangle M = 0$, $J = 0 \rightarrow 1$
transitions in the microwave region \cite{xx}  and $\triangle J =
0$, $|M| = 0 \rightarrow 1$ transitions in the radiofrequency region
\cite{Wharton} have been observed, in fields ranging from $\sim
200-500$ V/cm. For the radiofrequency transitions, line widths were
only about 2 kHz, consistent with just the dwell time in the
C-field. But for the microwave transitions the widths are much
larger, 45 kHz; this is attributed both to the higher frequency of
the transitions and to experimental conditions that render more
significant Doppler broadening and nonuniformity of the field,
especially in the entrance and exit fringe regions \cite{xw}. The
Doppler and dwell time contributions are not relevant to inferring
what might be expected for trapped BaO (or SrO). Broadening by
inhomogeneity of the external field is relevant but depends very
much on experimental particulars.  The transitions of interest,
depicted in Fig. \ref{Figure10}, occur in the microwave region and
involve Stark fields typically ten-fold larger than used in the
electric resonance spectroscopy, so the line widths might be
significantly broadened due to field inhomogeneity.    These
observations do not permit firm conclusions about the resolution
issue, but it decidedly poses an experimental challenge.

This discussion pertains only to the choice of qubits we have
considered, pendular states of linear polar molecules, which involve
transitions that change $\tilde{J}$ but not $M$.  The resolution
issue motivates examining other choices for qubits. For instance,
states with the same $\tilde{J}$ but different $M$ could be used.
Other options, particularly use of hyperfine or nuclear spin states
instead of  pendular states, have been suggested as means to reduce
sources of decoherence \cite{Demille,carr,YelinDeMille,andre}.  As
yet, the size of $\triangle\omega$ for any qubit choice other than
that used in this paper remains to be determined.

We intend to extend the treatment developed here to other choices
for qubit basis states as well as to larger numbers $N > 2$ of
dipoles. In preliminary work on linear and planar arrays of dipoles
up to $N = 8$, we find, as expected, the maximum pairwise
concurrence occurs for next-neighbor dipoles, although that for
non-nearest ones is significant. Also in prospect is an analogous
treatment of the proposed coupling of polar molecules via microwave
strip-lines \cite{andre}. There the entangling interaction differs
from direct dipole-dipole interactions, but again the proposed
qubits are pendular states.

Another pendular variant inviting attention is use of polar
symmetric top molecules.  The $|0\rangle$ and $|1\rangle$ qubits can
be selected as $|J,K,M\rangle = |1,1,-1\rangle$ and
$|1,1,+1\rangle$, which are degenerate in the field-free limit and
thus have a first-order Stark effect \cite{Townes}.  Even in a weak
electric field, these states are strongly oriented along and opposed
to the field, with equal and opposite projections. Moreover, the
effective dipole moments do not depend on the field strength so low
fields can be used if necessary to reduce line broadening, without
the penalty imposed by quenching of effective dipoles that would
occur for the second-order Stark effect. At first blush, the
symmetric top option appears to be disallowed because transitions
between the $M$ = -1 and +1 Stark states violate the selection
rules, $\triangle M = 0$ or $\triangle M = \pm 1$.  But the
prohibition is not absolute. Because the optical lattice perturbs
cylindrical symmetry about the field, $M$ is not strictly a "good"
quantum number, so the $\triangle M$ selection rule is relaxed.
Moreover, if the molecule contains an atom with nuclear spin $I >
1/2$, and hence an electric quadrupole moment, transitions with
$\triangle M = \pm 2$ become allowed. For instance, a deuterium
nucleius ($I = 1$) makes $\triangle M = \pm 2$ transitions facile in
Stark spectra \cite{Wofsy}. Other symmetric top options for qubits
are inversion doublets (e.g. in NH$_3$) or internal rotation states
associated with hindered torsional motion (e.g. CH$_3$CF$_3$); these
offer strong dipole-allowed transitions.

Previous studies of entanglement, both for polar molecules
\cite{yelin,Wei} and for magnetic spins
\cite{entg-kais,Osenda1,Osenda2,Osenda3,Osenda4,Xu}, have considered
primarily domains where the concurrence is large ($> 0.1$), and have
focused on means to tune the entanglement to attain such domains.
For polar molecules, that requires $\Omega/B > 1$. Recently, it was
suggested that such large $\Omega/B$ could be attained for dipole
arrays by exploiting nanotraps with lattice spacing of the order of
only 10 nm \cite{Wei}. However, as emphasized in Sec. IV, for
quantum computing large entanglement in the ground eigenstate is not
required.  Indeed, reducing the array spacing so markedly would
strongly foster inelastic, spontaneous Raman scattering of lattice
photons and hence induce decoherence
\cite{Demille,carr,YelinDeMille}. Such considerations make small
rather than large $\Omega/B$, and consequently weak rather than
strong entanglement in the ground eigenstate, actually preferable
for quantum computing \cite{Wootters}, provided resolution of the
$\triangle\omega$ shift can be attained.

\section*{ACKNOWLEDGEMENTS}
We are grateful to the Army Research Office for support of this work
at Purdue and to the National Science Foundation (CHE-0809651), the
Office of Naval Research and Hackerman Advanced Research program for
support at Texas A\&M University. We thank David DeMille and Seth
Lloyd for instructive discussions clarifying the role of
entanglement in quantum computing; Christopher Ticknor for calling
attention to kindred aspects of "dressing" dipole-dipole collisions
in an external electric field; and William Klemperer for elucidating
field effects in electric resonance spectroscopy and how to evade
the usual Stark transition selection rules.

\renewcommand{\theequation}{A\arabic{equation}}
\setcounter{equation}{0}  
\section*{APPENDIX A: DIPOLE-DIPOLE INTERACTION}

The angular dependence of the dipole-dipole interaction, given in
Eq.(\ref{coupling}), is usually expressed as
\begin{equation}
\Phi_{ij}=\text{cos}\beta-3\text{cos}\gamma_i\text{cos}\gamma_j
\end{equation}
where $\beta$ is the angle between dipoles $\mu_i$ and $\mu_j$;
angles $\gamma_i$ and $\gamma_j$ specify the orientation of the
dipoles with respect to the vector $r_{ij}$ between them.
Ordinarily, it is natural (and done in all textbooks) to express
$\text{cos}\beta$ in terms of the angles $\gamma$ together with the
azimuthal angles $\phi_r$ about the $r_{ij}$ axis. Thus, use
\begin{equation}
\text{cos}\beta=\text{cos}\gamma_i\text{cos}\gamma_j+\text{sin}\gamma_i\text{sin}\gamma_j\text{cos}(\phi_{ri}-\phi_{rj})
\end{equation}
which when combined with the
-3$\text{cos}\gamma_i\text{cos}\gamma_j$ term gives the familiar
expression \cite{Stone}. In the presence of the external electric
field, we need to  recast $\Phi_{ij}$ in terms of angles $\theta_i$
and $\theta_j$ that specify the orientation of the dipoles with
respect to the direction of the external electric field. Therefore,
we use
\begin{equation}
\text{cos}\beta=\text{cos}\theta_i\text{cos}\theta_j+\text{sin}\theta_i\text{sin}\theta_j\text{cos}(\phi_{Ei}-\phi_{Ej})
\end{equation}
\begin{equation}
\text{cos}\gamma=\text{cos}\theta_i\text{cos}\alpha+\text{sin}\theta_i\text{sin}\alpha\text{cos}(\phi_{Ei}-\phi_{Er})
\end{equation}
where the $\phi_E$ are azimuthal angles about the field vector {\bf
E} and $\alpha$ is the angle between the field vector and the
$r_{ij}$ vector. The azimuthal factors can be expressed as
\begin{equation}
\text{cos}(\phi_{Ei}-\phi_{Ej})=\text{cos}\phi_{Ei}\text{cos}\phi_{Ej}+\text{sin}\phi_{Ei}\text{sin}\phi_{Ej}
\end{equation}
As $M$ = 0 states, which do not depend on the $\phi$ angles, are
chosen as the qubit basis states, in evaluating matrix elements of
$\Phi_{ij}$ between these states the integrations over
$d\phi_id\phi_j$ (from 0 to 2$\pi$) eliminate all terms involving
the $\phi_E$ angles. The net result is simply
\begin{equation}
\langle\Phi_{ij}\rangle_\phi=(1-3\text{cos}^2\alpha)\text{cos}\theta_i\text{cos}\theta_j
\end{equation}
The effect of integrating over the $\theta$ angles is just to
replace in $V_{dd}$ the dipole moments $\mu_i$ and $\mu_j$ by their
effective values, $\mu\langle\text{cos}\theta\rangle$. The effective
dipole-dipole interaction hence reduces to
\begin{equation}
V_{d-d}=\Omega(1-3\text{cos}^2\alpha)\langle\text{cos}\theta_i\rangle\langle\text{cos}\theta_j\rangle
\end{equation}
with $\Omega=\mu_i\mu_j/r_{ij}^3$ as a convenient scale factor.

\renewcommand{\theequation}{B\arabic{equation}}
\setcounter{equation}{0}  

\section*{APPENDIX B: ZERO-FIELD CASE}

For $\mu${\Large $\mathbf{ \varepsilonup}$}/$B$ = 0, the Hamiltonian
matrix reduces to diagonal terms from Eq.(12) and antidiagonal
elements from Eq.(13), and the pendular qubit basis states become
simply $|0\rangle = Y_{00}$ and $|1\rangle = Y_{10}$. The form of
the Hamiltonian makes it equivalent to that for the Ising model for
a system with two qubits \cite{Peter}. Diagonalization of the
Hamiltonian yields the eigenenergies and eigenvectors given in Table
II as explicit functions of $\Omega/B$.

\begin{table}[H]
\caption{Zero-Field limit for N = 2 dipoles.} \centering
\begin{minipage}{11cm}
\begin{tabular}{cccc}
\hline \hline
\;\;\;\;$i$ &\;\;\;\;\;\;\; Eigenenergies, $E_i/B$ & \;\;\;\;\;\;\; Wavefunction, $\Psi_i$ \;\;\;\;\; & \;\;\;\;\;\;\; $C_{12}$ \;\;\;\;\;\\
\hline \let\thefootnote\relax\footnotetext{where $\alpha_{\pm}=
\left[1\pm\left(1+\zeta^2\right)^{1/2}\right]/\zeta$, with $\zeta =
\Omega/6B$}
\;\;\;\; 1 & \;\;\;\;\;\;\; $2-2(1+\zeta^2)^{1/2}$ & \;\;\;\;\;\;\; $\frac{1}{\sqrt{1+\alpha_+^2}}(|11\rangle-\alpha_+|00\rangle)$ \;\;\;\;\; & \;\;\;\;\;\;\;$\frac{2\alpha_+}{1+\alpha_+^2}$ \;\;\;\; \\
\;\;\;\; 2 & \;\;\;\;\;\;\; $2(1-\zeta)$  & \;\;\;\;\;\;\; $\frac{1}{\sqrt{2}}(|10\rangle-|01\rangle)$ \;\;\;\;\; & \;\;\;\;\;\;\; 1 \;\;\;\; \\
\;\;\;\; 3 & \;\;\;\;\;\;\; $2(1+\zeta)$ & \;\;\;\;\;\;\; $\frac{1}{\sqrt{2}}(|10\rangle+|01\rangle)$ \;\;\;\;\;  & \;\;\;\;\;\;\; 1 \;\;\;\; \\
\;\;\;\; 4 & \;\;\;\;\;\;\; $2+2(1+\zeta^2)^{1/2}$ & \;\;\;\;\;\;\; $\frac{1}{\sqrt{1+\alpha_-^2}}(|11\rangle-\alpha_-|00\rangle)$ \;\;\;\;\; & \;\;\;\;\;\;\;$\frac{2|\alpha_-|}{1+\alpha_-^2}$ \;\;\;\; \\
\hline
\end{tabular}
\end{minipage} \label{tablec1}
\end{table}

The density matrix for eigenstate 1, the ground-state, is
\begin{equation}
\rho(1) = |\Psi_1\rangle \langle\Psi_1| = \frac{1}{1+\alpha_+^2}
\left(\begin{array}{cccc}
\alpha_+^2  & 0 & 0 & -\alpha_+   \\
 0 & 0 & 0 & 0   \\
 0 & 0 & 0 & 0 \\
 -\alpha_+ & 0 & 0 & 1
\end{array} \right)
\end{equation}
That for eigenstate 2 is
\begin{equation}
\rho(2) = |\Psi_2\rangle \langle\Psi_2| = +\frac{1}{2}
\left(\begin{array}{cccc}
 0 & 0 & 0 & 0   \\
 0 & 1 & -1 & 0   \\
 0 & -1 & 1 & 0 \\
 0 & 0 & 0 & 0
\end{array} \right)
\end{equation}
The $\rho(3)$ matrix differs from $\rho(2)$ by having $+1$ in place
of each $-1$; the $\rho(4)$ matrix differs from $\rho(1)$ by having
$\alpha_-$ in place of $\alpha_+$. As these density matrices pertain
to only two dipoles, they need not be reduced further.

Obtaining the density matrices, $\tilde{\rho}(i)$, for the
spin-flipped states, defined in Eq. (\ref{rhotilde}), involves
shuffling the rows and columns of $\rho(i)$ in accord with
$|00\rangle \leftrightarrow |11\rangle$ and $|01\rangle
\leftrightarrow |10\rangle$. This gives
\begin{equation}
\tilde{\rho}(1) = \frac{1}{1+\alpha_+^2} \left(\begin{array}{cccc}
 1 & 0 & 0 & -\alpha_+   \\
 0 & 0 & 0 & 0   \\
 0 & 0 & 0 & 0 \\
 -\alpha_+ & 0 & 0 & \alpha_+^2
\end{array} \right)
\label{rhotilde1}
\end{equation}
and thus the product matrix is
\begin{equation}
\rho(1)\tilde{\rho}(1) = \frac{1}{\left(1+\alpha_+^2\right)^2}
\left(\begin{array}{cccc}
 2\alpha_+^2 & 0 & 0 & -\alpha_+^3   \\
 0 & 0 & 0 & 0   \\
 0 & 0 & 0 & 0 \\
 -2\alpha_+ & 0 & 0 & 2\alpha_+^2
\end{array} \right)
\label{rhorhotilde1}
\end{equation}
The eigenvalues of this matrix are $\lambda_1 =
4\alpha_+^2/\left(1+\alpha_+^2\right)^2$, $\lambda_2$ = $\lambda_3$
= $\lambda_4$ = 0. From Eq. (\ref{concurrence}), the concurrence is,
$C_{12}(1)$ = $2\alpha_+/(1+\alpha_+^2)$. Similarly, we find the
concurrences for the other eigenstates, given in Table I.

To evaluate the thermal concurrence, we need to set up the thermal
density matrix
\begin{equation}
\rho(T) =
\sum_{i=1}^4\text{exp}\left(-E_i/k_BT\right)|\Psi_i\rangle\langle\Psi_i|
= \left(\begin{array}{cccc}
 a & 0 & 0 & g   \\
 0 & b & d & 0   \\
 0 & d & b & 0 \\
 g & 0 & 0 & c
\end{array} \right)
\label{thermalrho}
\end{equation}
where
\begin{equation}
a =
\frac{\alpha_+^2}{1+\alpha_+^2}P_1+\frac{\alpha_-^2}{1+\alpha_-^2}P_4
\end{equation}
\begin{equation}
b = \frac{1}{2}(P_2 + P_3)
\end{equation}
\begin{equation}
c = \frac{1}{1+\alpha_+^2}P_1+\frac{1}{1+\alpha_-^2}P_4
\end{equation}
\begin{equation}
d = \frac{1}{2}(P_3 - P_2)
\end{equation}
\begin{equation}
g =
-\frac{\alpha_+}{1+\alpha_+^2}P_1-\frac{\alpha_-}{1+\alpha_-^2}P_4
\end{equation}
with
\begin{equation}
P_i = \frac{\text{exp}(-E_i/k_BT)}{Z}
\end{equation}

Then we find
\begin{equation}
\rho(T)\tilde{\rho}(T) = \left(\begin{array}{cccc}
 ac+g^2 & 0 & 0 & 2ag   \\
 0 & b^2+d^2 & 2bd & 0   \\
 0 & 2bd & b^2+d^2 & 0 \\
 2cg & 0 & 0 & ac+g^2
\end{array} \right)
\label{rhorhotildeT}
\end{equation}
and obtain the eigenvalues,
\begin{equation}
\lambda_1 = (\sqrt{ac}-g)^2,\;\;\;\lambda_2 =
(b-d)^2,\;\;\;\lambda_3 = (b+d)^2,\;\;\;\lambda_4 = (\sqrt{ac}+g)^2
\end{equation}
Hence from Eq. (\ref{concurrence}), we obtain the thermal
concurrence,
\begin{equation}
C_{12}(T) = max\left\{0, -2(b+g) \right\} \label{Ising-C12}
\end{equation}
With the $C_{12}(i)$ of Table I, this gives Eq.(20) of the text,
\begin{equation}
C_{12}(T)=C_{12}(1)P_1-C_{12}(2)P_2-C_{12}(3)P_3-C_{12}(4)P_4
\end{equation}

When $\Omega/B << 1$, the ground state concurrence becomes
$C_{12}(1)\rightarrow \Omega/6B$, in accord with Eq.(\ref{C12}) of
the text, whereas $C_{12}(4)\rightarrow \Omega/6B$ and
$C_{12}(2)=C_{12}(3)=1$. Provided that also $k_BT/B << 1$, a
first-order expansion of Eq.(B15) gives
\begin{equation}
C_{12}(T) \approx C_{12}(1)-\epsilon_2-\epsilon_3,
\end{equation}
where $\epsilon_2 = P_2/P_1 << 1$; $\epsilon_3 = P_3/P_2 << 1$ and
$P_4/P_1 <<< 1$. Then
\begin{equation}
C_{12}(T) \approx K(0)\left[\Omega/B - \Omega_c/B\right],
\end{equation}
which has the form of Eq.(21) of the text with $K(0)=1/6$ and
\begin{equation}
y_0(T) = \Omega_c/B = (\epsilon_2+\epsilon_3)/K(0) =
12\text{exp}\left(-2B/k_BT\right)\text{cosh}\left(\Omega/3k_BT\right)
\end{equation}

This result for the zero-field case, although not useful in
practice, illustrates how the excited states are involved in
creating a temperature dependent minimum level of dipole-dipole
coupling, $\Omega_c/B$, that is required to have nonzero thermal
concurrence.

Figure \ref{Figure13} shows a contour plot of $C_{12}(T)$ for the
zero-field case, derived from Eq.(B14).  It is qualitatively quite
similar to Fig. 8 for the pendular case, over the same wide range of
$k_BT/B$ and $\Omega/B$.

\renewcommand{\theequation}{C\arabic{equation}}
\setcounter{equation}{0}  
\section*{APPENDIX C: REDUCED VARIABLE FORMULAS}
In order to find a proper reduced variable formula, three steps are
needed: (1) calculate enough sample points to define well the exact
curve; (2) find a function with adjustable parameters that enables
fitting those points; (3) evaluate the parameters using a non-linear
regression method. For our curve fitting we use the
Levenberg-Marquardt Algorithm \cite{Kenneth}, also called
"Chi-square minimization". Chi-square is defined as:
\begin{equation}
\chi^2 = \sum_{i=1}^N \left[Y_i - f(x_i;\hat{\theta})\right]^2
\end{equation}
where $x_i$ and $Y_i$ are the independent and dependent variables
for the $i^{th}$ (i = 1, 2,...,n) sample points of the exact curve;
$\hat{\theta}$ are the parameters to be fitted. The
Levenberg-Marquardt algorithm iteratively adjusts the parameters to
get the minimum chi-square value, which corresponds to the best fit.
The input data for fitting Eqs.($19,22,25,26$) comprised our
numerical results for the pendular case, over the ranges $x =
\mu${\Large $\mathbf{ \varepsilonup}$}/$B$ = 0 to 8. Tables III - VI
list the optimal values found for the parameters and 95\% percent
confidence intervals. At $x = 0$, the Eq.(19) fit gives $K(0) =
0.17103$, different slightly from the exact zero-field limit, $K(0)
= 1/6$. Likewise, at $x=0$ the Eq.(25) and (26) fits give $C_0 =
0.005$ and $C_1 = 0.00072$ rather than the exact value of zero. The
critical point for $C_1$ to change sign is $x=4.902$, whereas
Eq.(26) gives $C_1 = -0.00025$, slightly different from zero.

\begin{table}[H]
\caption{Values of the parameters for Eq.(\ref{kc}).}\centering
\begin{minipage}{11cm}
\begin{tabular}{cccc}
\hline\hline \;\;\;\;\;\;Parameters \;\;\;\;\;\;& \;\;\;\; Pendular
\;\;\;\;\;\;& \;\;\;\; Field-free \;\;\;\;\;\; & \;\;\;\;
CI\footnote{95\% confidence interval; values listed are maximum
found for the 2
curves shown in Fig.\ref{Figure6}. Both $R^2$ values are around 0.9981.Similarly accurate results are found when Eq.(\ref{C12}) is generalized for different E-fields at the dipole sites by replacing $K(x)$ by $[K(x)K(x')]^{1/2}$.} \;\;\;\; \\
\hline
 \;\;\;\;\;\;$A_1$ \;\;\;\;\;\; & \;\;\;\; 0.01092 \;\;\;\;\;\; & \;\;\;\; 0.00221 \;\;\;\;\;\; & \;\;\;\; $\pm$0.0003 \;\;\;\; \\
 \;\;\;\;\;\;$A_2$ \;\;\;\;\;\; & \;\;\;\; 0.21953 \;\;\;\;\;\; & \;\;\;\; 0.24779 \;\;\;\;\;\; & \;\;\;\; $\pm$0.006 \;\;\;\; \\
 \;\;\;\;\;\;$x_0$ \;\;\;\;\;\; & \;\;\;\; 0.96578 \;\;\;\;\;\; & \;\;\;\; 0.74035 \;\;\;\;\;\; & \;\;\;\; $\pm$0.05 \;\;\;\; \\
 \;\;\;\;\;\;$\triangle x$ \;\;\;\;\;\; & \;\;\;\; 0.97429 \;\;\;\;\;\; & \;\;\;\; 0.86072 \;\;\;\;\;\; & \;\;\;\; $\pm$0.03 \;\;\;\; \\
\hline
\end{tabular}
\end{minipage}
\label{table1}
\end{table}

\begin{table}[H]
\centering \caption{Values of the parameters for Eq.(\ref{eqW1-W0}).
}
\begin{minipage}{9cm}
\begin{tabular}{ccc}
\hline\hline
Parameters \;\;\;\;\;\;\;\;\;\; & \;\;\;\;\; Values \;\;\;\;\;\;\;\;\;\;\;\;\; & \;\;\; CI\footnote{95\% confidence interval. $R^2 = 0.9999$} \;\;\; \\
\hline
 $A_1$ \;\;\;\;\;\;\;\;\;\; & \;\;\;\;\; 12.42379 \;\;\;\;\;\;\;\;\;\;\;\;\; & \;\;\;$\pm$0.0533 \;\;\; \\
 $A_2$ \;\;\;\;\;\;\;\;\;\; & \;\;\;\;\; -10.47646 \;\;\;\;\;\;\;\;\;\;\;\;\; & \;\;\;$\pm$0.0534 \;\;\; \\
 $x_0$ \;\;\;\;\;\;\;\;\;\; & \;\;\;\;\; 8.77516 \;\;\;\;\;\;\;\;\;\;\;\;\; & \;\;\;$\pm$0.0534 \;\;\; \\
 $p$ \;\;\;\;\;\;\;\;\;\; & \;\;\;\;\; 1.5867 \;\;\;\;\;\;\;\;\;\;\;\;\; & \;\;\;$\pm$0.00527 \;\;\; \\
\hline
\end{tabular}
\end{minipage}
\label{tableW1-W0}
\end{table}

\begin{table}[H]
\centering \caption{Values of the parameters for Eq.(\ref{eqC0}). }
\begin{minipage}{9cm}
\begin{tabular}{ccc}
\hline\hline
Parameters \;\;\;\;\;\;\;\;\;\; & \;\;\;\;\; Values \;\;\;\;\;\;\;\;\;\;\;\;\; & \;\;\; CI\footnote{95\% confidence interval. $R^2 = 0.99994$} \;\;\; \\
\hline
 $A_1$ \;\;\;\;\;\;\;\;\;\; & \;\;\;\;\; 0.84855 \;\;\;\;\;\;\;\;\;\;\;\;\; & \;\;\;$\pm$0.00145 \;\;\; \\
 $A_2$ \;\;\;\;\;\;\;\;\;\; & \;\;\;\;\; -0.84355 \;\;\;\;\;\;\;\;\;\;\;\;\; & \;\;\;$\pm$0.00180 \;\;\; \\
 $x_0$ \;\;\;\;\;\;\;\;\;\; & \;\;\;\;\; 1.6339 \;\;\;\;\;\;\;\;\;\;\;\;\; & \;\;\;$\pm$0.00508 \;\;\; \\
 $p$ \;\;\;\;\;\;\;\;\;\; & \;\;\;\;\; 1.2459 \;\;\;\;\;\;\;\;\;\;\;\;\; & \;\;\;$\pm$0.00539 \;\;\; \\
\hline
\end{tabular}
\end{minipage}
\label{tableC0}
\end{table}

\begin{table}[H]
\centering \caption{Values of the parameters for Eq.(\ref{eqC1}). }
\begin{minipage}{9cm}
\begin{tabular}{ccc}
\hline\hline
Parameters \;\;\;\;\;\;\;\;\;\; & \;\;\;\;\; Values \;\;\;\;\;\;\;\;\;\;\;\;\; & \;\;\; CI\footnote{95\% confidence interval. $R^2 = 1$} \;\;\; \\
\hline
 $A_0$ \;\;\;\;\;\;\;\;\;\; & \;\;\;\;\; -0.75212 \;\;\;\;\;\;\;\;\;\;\;\;\; & \;\;\;$\pm$0.0323 \;\;\; \\
 $A_1$ \;\;\;\;\;\;\;\;\;\; & \;\;\;\;\; 1.04192 \;\;\;\;\;\;\;\;\;\;\;\;\; & \;\;\;$\pm$0.0336 \;\;\; \\
 $A_2$ \;\;\;\;\;\;\;\;\;\; & \;\;\;\;\; 1.14092 \;\;\;\;\;\;\;\;\;\;\;\;\; & \;\;\;$\pm$0.0325 \;\;\; \\
 $x_1$ \;\;\;\;\;\;\;\;\;\; & \;\;\;\;\; -0.16241 \;\;\;\;\;\;\;\;\;\;\;\;\; & \;\;\;$\pm$0.0224 \;\;\; \\
 $x_2$ \;\;\;\;\;\;\;\;\;\; & \;\;\;\;\; 3.1232 \;\;\;\;\;\;\;\;\;\;\;\;\; & \;\;\;$\pm$0.124 \;\;\; \\
 $\triangle x_1$ \;\;\;\;\;\;\;\;\;\; & \;\;\;\;\; 0.90544 \;\;\;\;\;\;\;\;\;\;\;\;\; & \;\;\;$\pm$0.0136 \;\;\; \\
 $\triangle x_2$ \;\;\;\;\;\;\;\;\;\; & \;\;\;\;\; 2.76286 \;\;\;\;\;\;\;\;\;\;\;\;\; & \;\;\;$\pm$0.0496 \;\;\; \\
\hline
\end{tabular}
\end{minipage}
\label{tableC1}
\end{table}

For convenience, we give formulas for the three unitless ratios,
evaluated with customary units:

$\mu${\Large $\varepsilonup$}$/B$ = 0.0168 $\mu(Debye)${\Large
$\varepsilonup$}$(kV/cm)/B(cm^{-1})$;

$\Omega/B = 5.04 \times 10^{-9} \mu^2
(Debye)/r^3(microns)/B(cm^{-1})$;

$k_BT/B = 0.695\;T(K)/B(cm^{-1})$.


\newpage

\newpage

\newpage

\newpage

\newpage
\clearpage

\newpage
\begin{figure}[t]
\begin{center}
\includegraphics[width=0.80\textwidth]{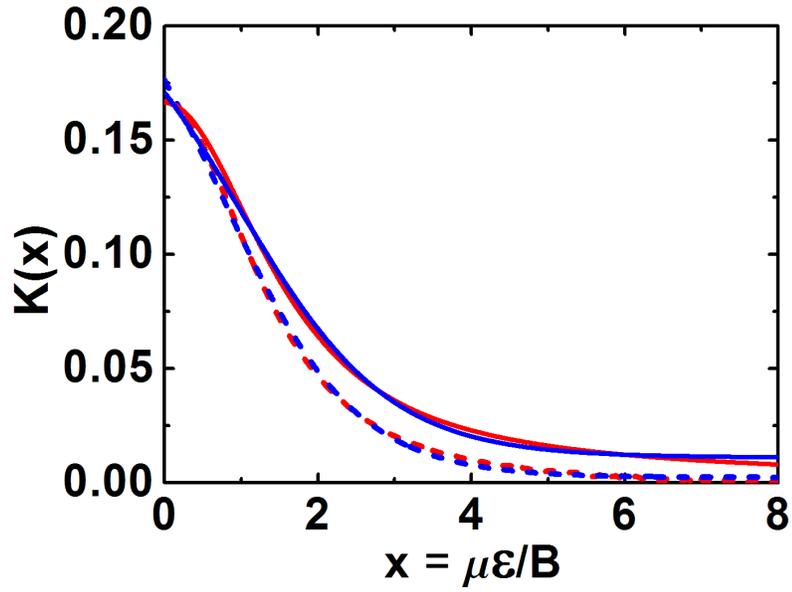}
\end{center}
\caption{(Color online) The $K(x)$ function of Eq.(18); solid curves
show exact result (red) and fitted function (blue) of Eq.(19) that
pertains to pendular qubit basis, see Eq.(3). For comparison, dashed
curve pertains to field-free basis with $|0\rangle$ = $Y_{0,0}$,
$|1\rangle$ = $Y_{1,0}$ ({\it Cf.} Table III, Appendix C).}
\label{Figure6}
\end{figure}

\newpage
\begin{figure}[t]
\begin{center}
\includegraphics[width=1.00\textwidth]{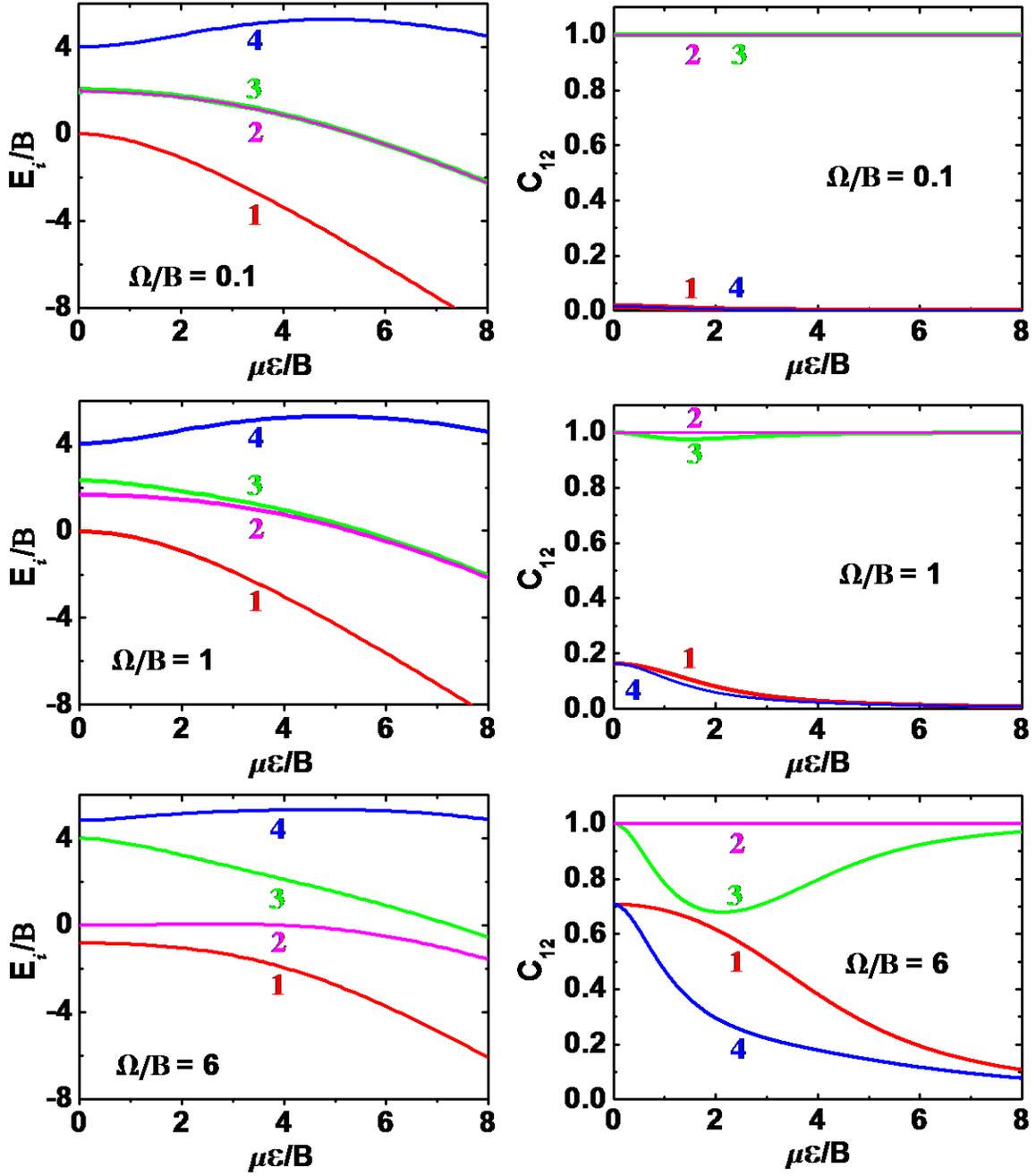}
\end{center}
\caption{(Color online) Eigenenergies and concurrences for the four
eigenstates for two dipoles as a function of reduced variables,
$\mu${\Large $\varepsilonup$}$/B$ for electric field and $\Omega/B$
for dipole-dipole coupling.} \label{Figure7}
\end{figure}

\newpage
\begin{figure}[t]
\begin{center}
\includegraphics[width=0.70\textwidth]{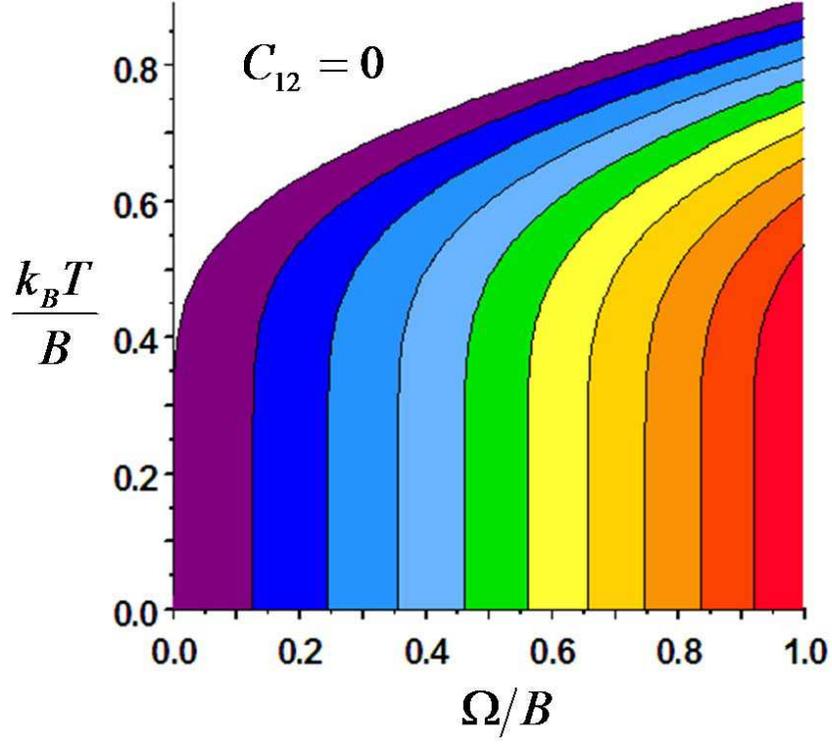}
\end{center}
\caption{(Color online) Contour plot of thermal pairwise concurrence
for two dipoles, for $\mu${\Large $\varepsilonup$}$/B$ = 3. For $0 <
\Omega/B < 1$, the maximum concurrence $C_{12}(max)$ = 0.0473,
occurs at T = 0, $\Omega/B$ = 1. The plot displays normalized
contours. Within each colored band, the variation of
$C_{12}/C_{12}(max)$ is 0.1; thus the normalized concurrence in the
right most band (red) ranges from 0.9 to 1; in the next band (orange
red), from 0.8 to 0.9, etc. A striking feature is the large region
(uncolored) where $C_{12}$ = 0. There, entanglement does not occur
unless the dipole-dipole coupling exceeds a critical value dependent
on the temperature.} \label{Figure8}
\end{figure}

\newpage
\begin{figure}[t]
\begin{center}
\includegraphics[width=0.70\textwidth]{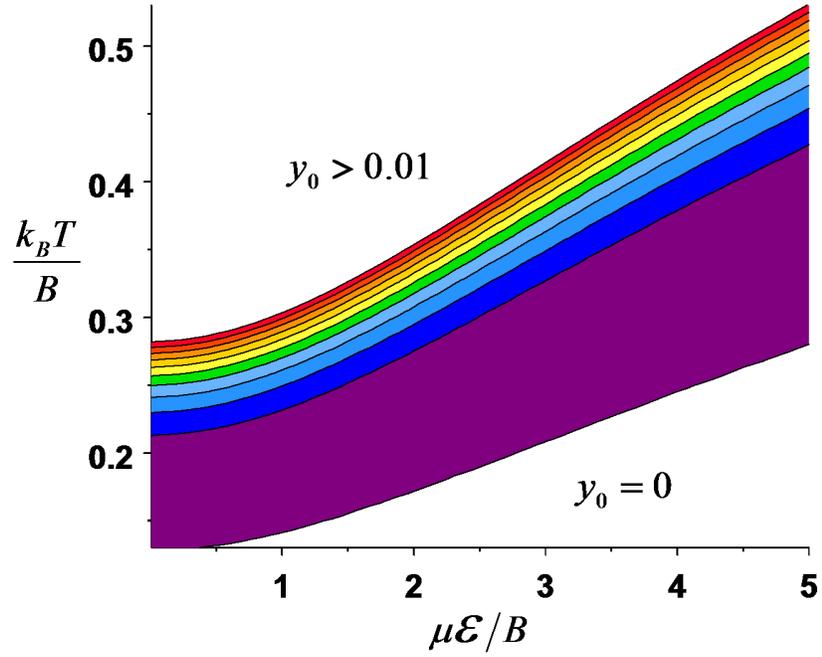}
\end{center}
\caption{(Color online) Contour plot displaying $y_0(x,z)$ term in
Eq.(21) vs. $x$ = $\mu${\Large $\varepsilonup$}$/B$. Within each
colored band, the range of $y_0$ is 0.001; thus in the lowest
colored band (magenta). $y_0$ ranges between 0 and 0.001; in the
highest colored band (red), $y_0$ is between 0.009 and 0.01.}
\label{Figure9}
\end{figure}

\newpage
\begin{figure}[t]
\begin{center}
\includegraphics[width=0.60\textwidth]{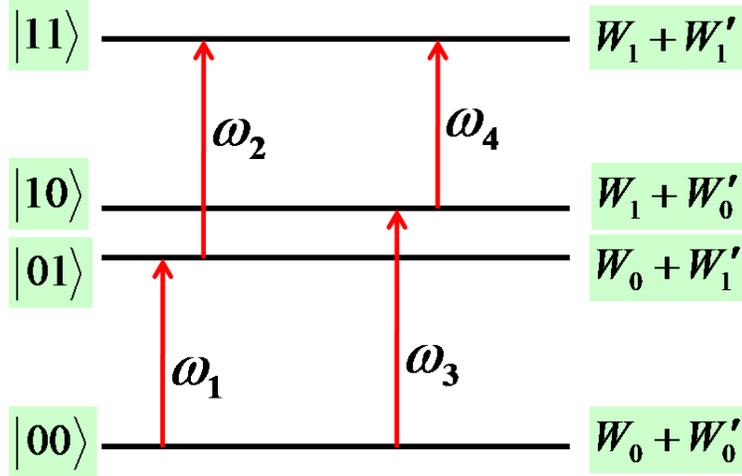}
\end{center}
\caption{(Color online) Schematic energy levels for qubit pendular
eigenstates of N = 2 dipoles, in absence of dipole-dipole
interaction, thus corresponding to Eq.(12).  Qubit basis states
shown at left, eigenenergies at right.  Pairs of transitions
involved in CNOT operation are indicated: $\omega_1$ transfers
dipole 2 from $|0\rangle$ to $|1\rangle$ with dipole 1 remaining in
$|0\rangle$; then $\omega_2$ transfers dipole 1 from $|0\rangle$ to
$|1\rangle$ with dipole 2 remaining in $|1\rangle$. Analogously, the
same result could be reached by $\omega_3$ followed by $\omega_4$.
Transition energies (including dipole-dipole terms to first-order)
are given in Eqs.(23).} \label{Figure10}
\end{figure}

\newpage
\begin{figure}[t]
\begin{center}
\includegraphics[width=0.60\textwidth]{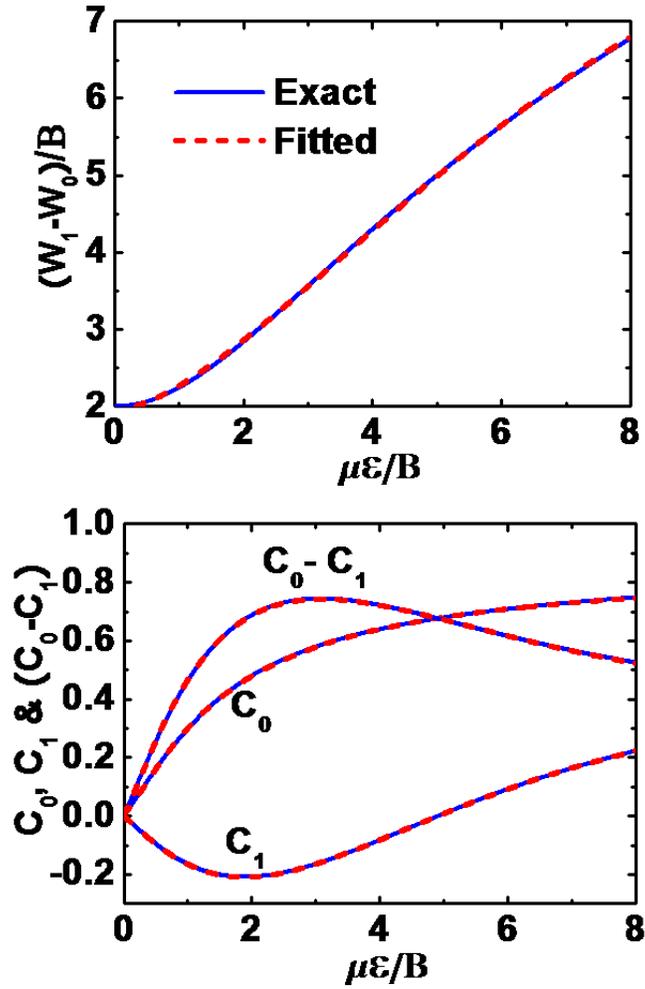}
\end{center}
\caption{(Color online) Comparison of exact results (blue curves)
with fitted approximation functions (dashed red curves) for
properties governing transitions among qubit states, Eqs.(23):
pendular energy difference, $(W_1 - W_0)/B$, cosine expectation
values, $C_0$ and $C_1$ and their difference, $C_0 - C_1$; {\it cf.}
Eqs. (\ref{eqW1-W0}, \ref{eqC0}, \ref{eqC1}).} \label{Figure11}
\end{figure}

\newpage
\begin{figure}
\begin{center}
\includegraphics[width=1.0\textwidth]{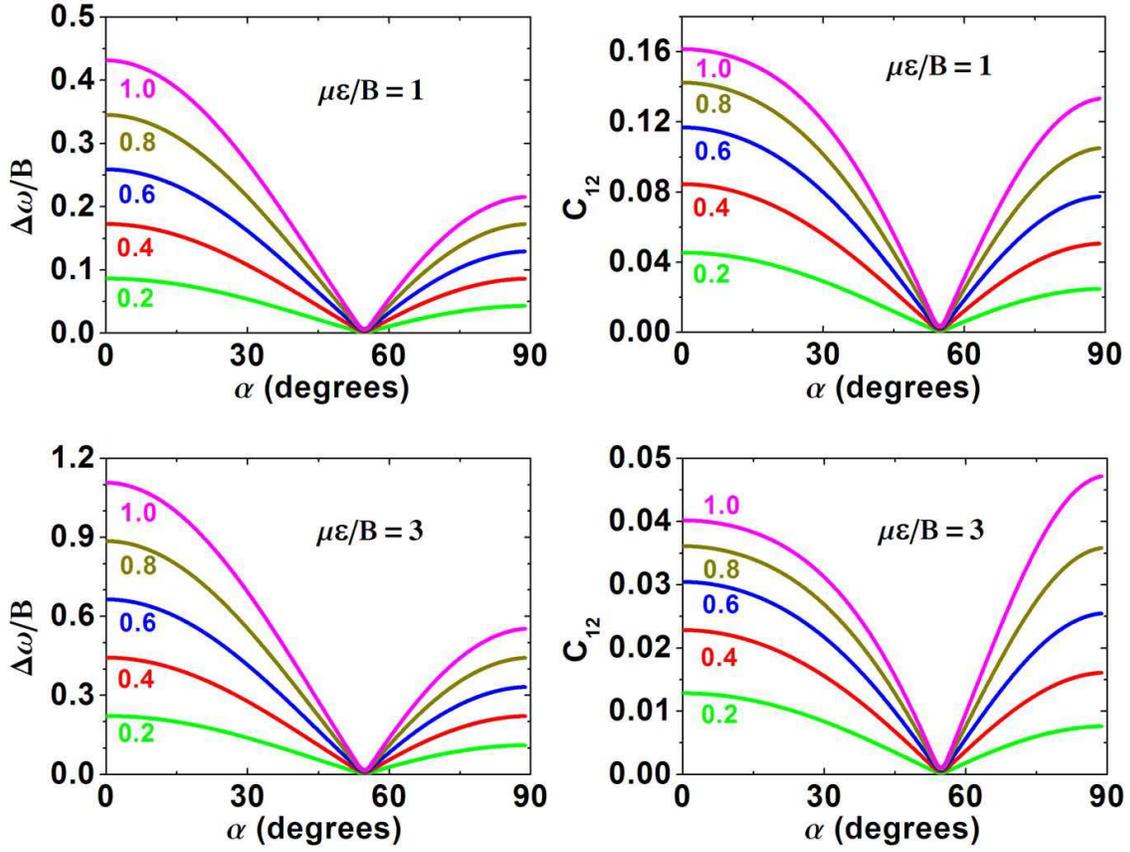}
\end{center}
\caption{(Color online) Frequency shift $\triangle\omega/B$ (left
panels) and ground state concurrence $C_{12}$ (right panels) as
functions of $\alpha$, the orientation angle of the electric field.
Curves are shown for $\Omega/B$ = 0.2 to 1.0 with $\mu${\Large
$\varepsilonup$}/$B$ = 1 or 3.} \label{Figure12}
\end{figure}

\newpage
\begin{figure}[t]
\begin{center}
\includegraphics[width=0.70\textwidth]{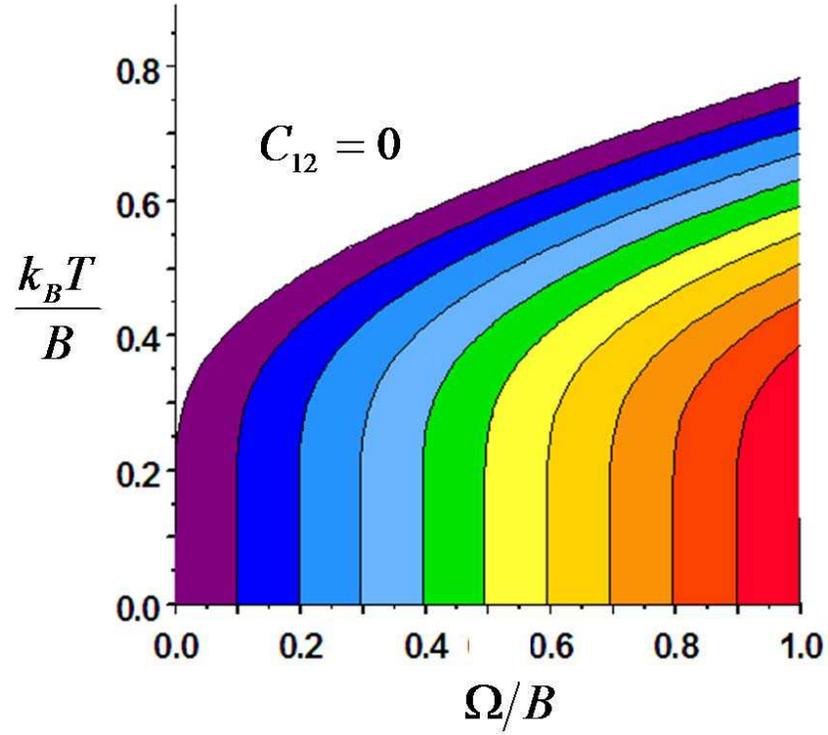}
\end{center}
\caption{(Color online) Contour plot of thermal pairwise concurrence
for field-free case, prepared in same format for comparison with
Fig. 8 for the pendular case. Here, $C_{12}(max)$ = 0.1644 at T = 0,
$\Omega/B$ = 1.} \label{Figure13}
\end{figure}


\begin{thebibliography}{51}
\bibitem{Thanks} Our discussion of the CNOT operation is largely drawn from tutorial instruction kindly given us by D. DeMille, amplifying the discussion of Fig.17.2 in ref. 13.
\bibitem{xw} Our discussion of line widths is based on conversations with W. Klemperer, interpreting observations made in refs. \cite{xx} and \cite{Wharton}.
\end{thebibliography}
\end{document}